\documentclass[
 amsmath,amssymb,
aps,prb,reprint,
]{revtex4-2}
\usepackage{graphicx}
\usepackage{dcolumn}
\usepackage{bm}
\usepackage{placeins}
\usepackage{appendix}
\usepackage[utf8]{inputenc}
\usepackage[english]{babel}
\usepackage{float}

\newcommand{ \angstrom}{\mbox{\normalfont\AA}}

\begin{document}
\UseRawInputEncoding

\title[CSPEC: The cold chopper spectrometer of the European Spallation Source, a detailed overview prior to commissioning.]{CSPEC: The cold chopper spectrometer of the European Spallation Source, a detailed overview prior to commissioning.}

\author{P.P.Deen}
\affiliation{European Spallation Source, Partikelgatan, 224 84 Lund, Sweden}
\affiliation{Nanoscience Center, Niels Bohr Institute, University of Copenhagen, 2100 Copenhagen Ø, Denmark}
\email{pascale.deen@ess.eu}

\author{F. Moreira}%
\affiliation{European Spallation Source, Partikelgatan, 224 84 Lund, Sweden}

\author{D. Noferini}%
\affiliation{European Spallation Source, Partikelgatan, 224 84 Lund, Sweden}

\author{S. Longeville}
\affiliation{%
Laboratoire L{\'e}on Brillouin, Université de Paris-Saclay, CEA/CNRS UMR 12, CEA-Saclay, 91191 Gif-sur-Yvette cedex,France
}%

\author{G. Fabr{\'e}ges}%
 \affiliation{%
Laboratoire L{\'e}on Brillouin, Université de Paris-Saclay, CEA/CNRS UMR 12, CEA-Saclay, 91191 Gif-sur-Yvette cedex,France
}%

\author{W.Lohstroh}%
\affiliation{ 
Heinz Maier-Leibnitz Zentrum, Technische Universität M{\"u}nchen, 85748 Garching, Germany
}%

\author{L.Loaiza}%
 \affiliation{ 
Heinz Maier-Leibnitz Zentrum, Technische Universität M{\"u}nchen, 85748 Garching, Germany
}%

\date{\today}

\begin{abstract}
CSPEC is the cold chopper spectrometer of the European Spallation Source (ESS) and will come on-line with the ESS beam on target. CSPEC will be the first cold chopper spectrometer on a long pulsed spallation source which provides great opportunities in terms of signal to noise and novel measuring schemes. We provide a detailed overview of the instrument, scientific design considerations and engineering requirements. 
\end{abstract}

\maketitle
\section{Measuring dynamics in materials}
The properties and functionality of materials are derived from where atoms are and how they move. Understanding atomic positions and movement is therefore essential to test, in a stringent manner, theoretical models of materials. Dynamic behaviour in materials can vary enormously. Collective modes such as phonon and magnons, with picosecond timescales, can determine thermal and electrical conductivity of materials \cite{PhysRevMaterials.3.025403_Phonons_Magnons, PhysRevB.99.174437Fe2P}. Quantum fluctuations give rise to new states of matter in the nanosecond to picosecond regime \cite{Gingras_QuantumFluctuctuations}. Rotational and translational diffusion are important for many biophysical processes, in the pico to microsecond regime \cite{SMITH_Proteins, HAMANEH_Proteins}. A chopper spectrometer can access a broad range of timescales from picosecond to nanosecond fluctuations, with the associative spatial scales, and is thus a work horse instrument for understanding material properties. In this paper we provide the scientific and technical requirements that have led to the design and construction of the cold chopper spectrometer of the ESS, CSPEC.  

  Inelastic neutron scattering encompasses broad energy and spatial domains and is generally subdivided into two distinct regimes. The first, a measure of the energy transfer of neutrons interacting with matter with scattering processes at small energy scales, quasielastic neutron scattering (QENS). Second, generally at higher energy transfers, through the creation and annihilation of elementary excitations, inelastic neutron scattering (INS). The resultant double partial differential cross section, for both QENS and INS, is proportional to the dynamic structure factor S($\textbf{Q}$, $\omega$), and depends on the wavevector transfer between sample and neutrons, $\textbf{Q}$, and the neutron energy transfer between sample and neutrons, $\omega$, respectively. 
  
  Neutron scattering is particularly effective since S($\textbf{Q}$, $\omega$) can be derived within the Born approximation which can be expressed directly in terms of space and time dependent correlations. Indeed S($\textbf{Q}$, $\omega$) represents the space and time Fourier Transform of the probability of finding atoms or spins, $i$,  separated by a particular distance to atom or spin, $j$, at a particular time t, r$_{i,j}$(t). As such, one directly compares the experimental signatures, S($\textbf{Q}$, $\omega$),  with complex theoretical models of the material probed \cite{Boothroyd2020}. 
  
 A direct geometry spectrometer is a work horse instrument for any neutron scattering research facility providing an unparalleled overview of S($\textbf{Q}$, $\omega$) with simultaneous insight into spatial and dynamic correlations over of a wide region of reciprocal space \cite{OllivierIN5, Lechner1990}. Direct geometry chopper spectrometers exist at reactor sources, for instance IN5 (ILL) \cite{IN5_ILL}, TOFTOF (FRM2) \cite{TOFTOF}   and DCS (NIST) \cite{DCS} and short pulse spallation sources such as LET (ISIS) \cite{BEWLEY_LET}, CNCS \cite{EHLERS_CNCS} (SNS) and AMATERAS \cite{AMATERAS_Kenji} (J-Parc). Here, cold chopper spectrometers are singled out to make a direct relevant comparison with CSPEC. Each instrument aims to optimise signal to noise within a rate that permits the scattering, from the sample, of neutrons with nearly infinite energy gain to those with energy losses down to 20\% of the incident neutron energy. A reactor source instrument is usually designed to optimise flux for a single incident wavelength while a spallation source instrument is able to provide a broad incident wavelength band via the use of repetition rate multiplication (RRM). To date, the relative flux on sample between reactor and spallation source chopper spectrometers depends on the peak brilliance of the facility and the ratio of neutron pulse on sample (50 - 300 Hz, typically), on a reactor source, to the frequency of the spallation source (25 - 60 Hz). Adjacent pulses, within a single spallation period, cannot be accumulated due to their widely varying incident energies on these instruments. In contrast, on CSPEC, scattered neutrons with closely matched incident energies can be accumulated to improve flux on sample. The noise at the various facilities can vary significantly. A reactor source instrument is always affected by the continuous operation of the reactor, providing a time independent neutron background, while a spallation source instrument will observe a burst of high energy neutrons from the prompt pulse of the spallation process but is clean in between. 
 
 CSPEC, the cold chopper spectrometer of the ESS, is optimised to harness the significant peak and average flux of the European Spallation Source \cite{ANDERSEN_ESS_Paper} while taking advantage of the low noise levels that accompany spallation. CSPEC has been developed to study life sciences, energy and functional materials in addition to emergent magnetism. The weak interaction of neutrons with matter, in addition to the limited flux at present-day neutron sources, make it currently difficult to address pertinent scientific questions such as those pertaining to materials with limited synthesis capabilities or accessing operando behaviour. The enhanced flux of the ESS, in addition to an optimised use of the ESS pulse structure on CSPEC while reducing noise, via careful consideration of the guide parameters, will open new scientific paths in materials research.

\section{Overview of time of flight spectroscopy at a spallation source.}

The momentum, $p$, and kinetic energy, $E$, of neutrons are linked, according to quantum theory, to the wavelength, $\lambda$, of the neutron via the de Broglie relation, which for non-relativistic matter, such as cold or thermal neutrons is given by,
\begin{equation}
E = \frac{p^2}{2m_n}= \frac{\hbar^2k^2}{2m_n} , 
\end{equation}
where $\hbar$ is the reduced Plank's constant, $k = \frac{2\pi}{\lambda}$  and $m_n$ = 1.675 x 10$^{-27}$ kg is the neutron mass.    
The CSPEC guide extracts neutrons from the ESS 20 K parahydrogen moderator. The kinetic energy of the neutrons equals 1.72 meV for T = 20 K given by $k_B T$ with  $k_B = 1.381 x 10^{-23}$ JK$^{-1}$. This energy, via de Broglie relation, provides a frequency of 0.42 THz directly relevant to the kinetic energies and fluctuations found in materials.  

\begin{figure}[h!]
	\centering
	\includegraphics[width=0.5\textwidth]{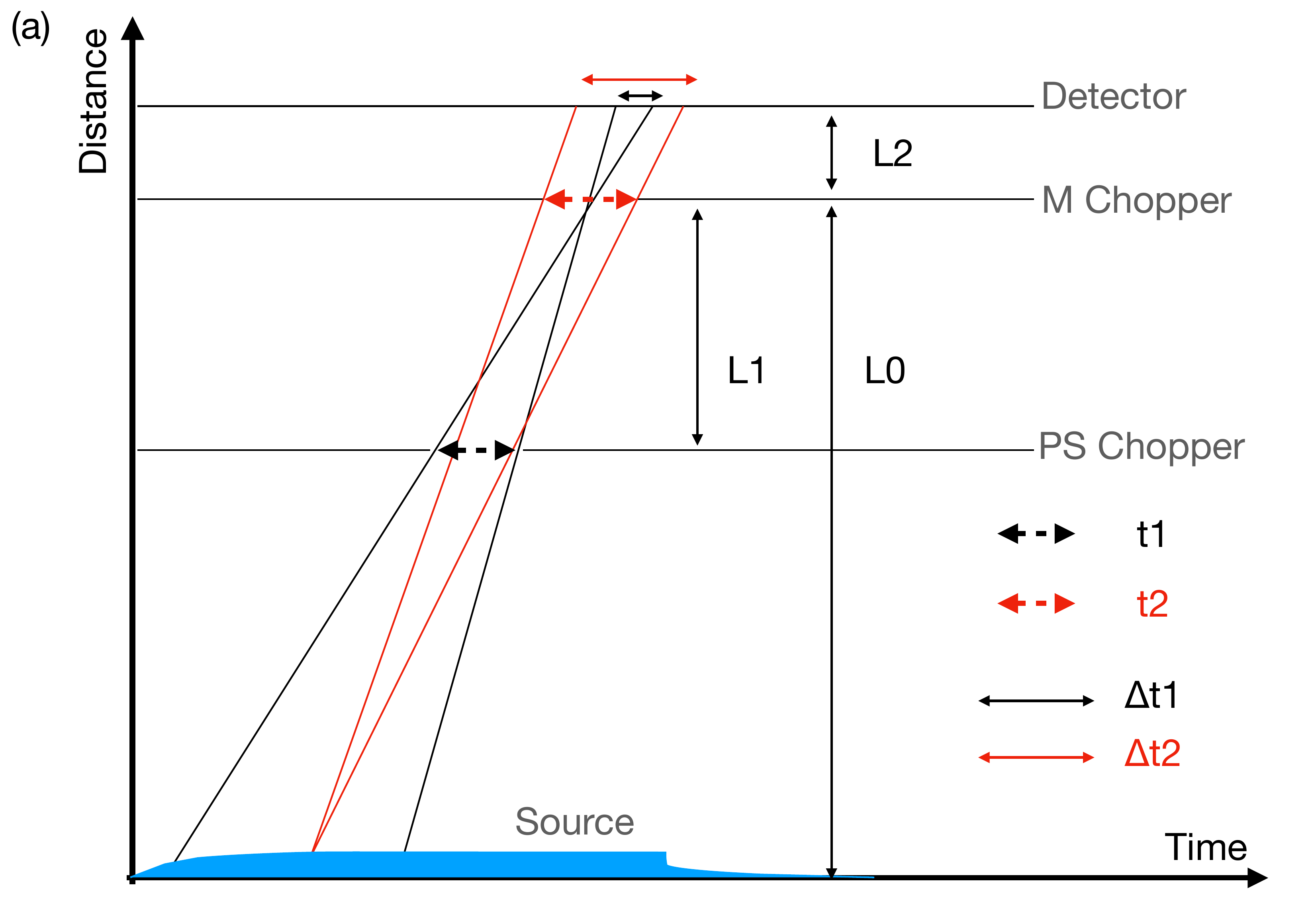}
\includegraphics[width=0.5\textwidth]{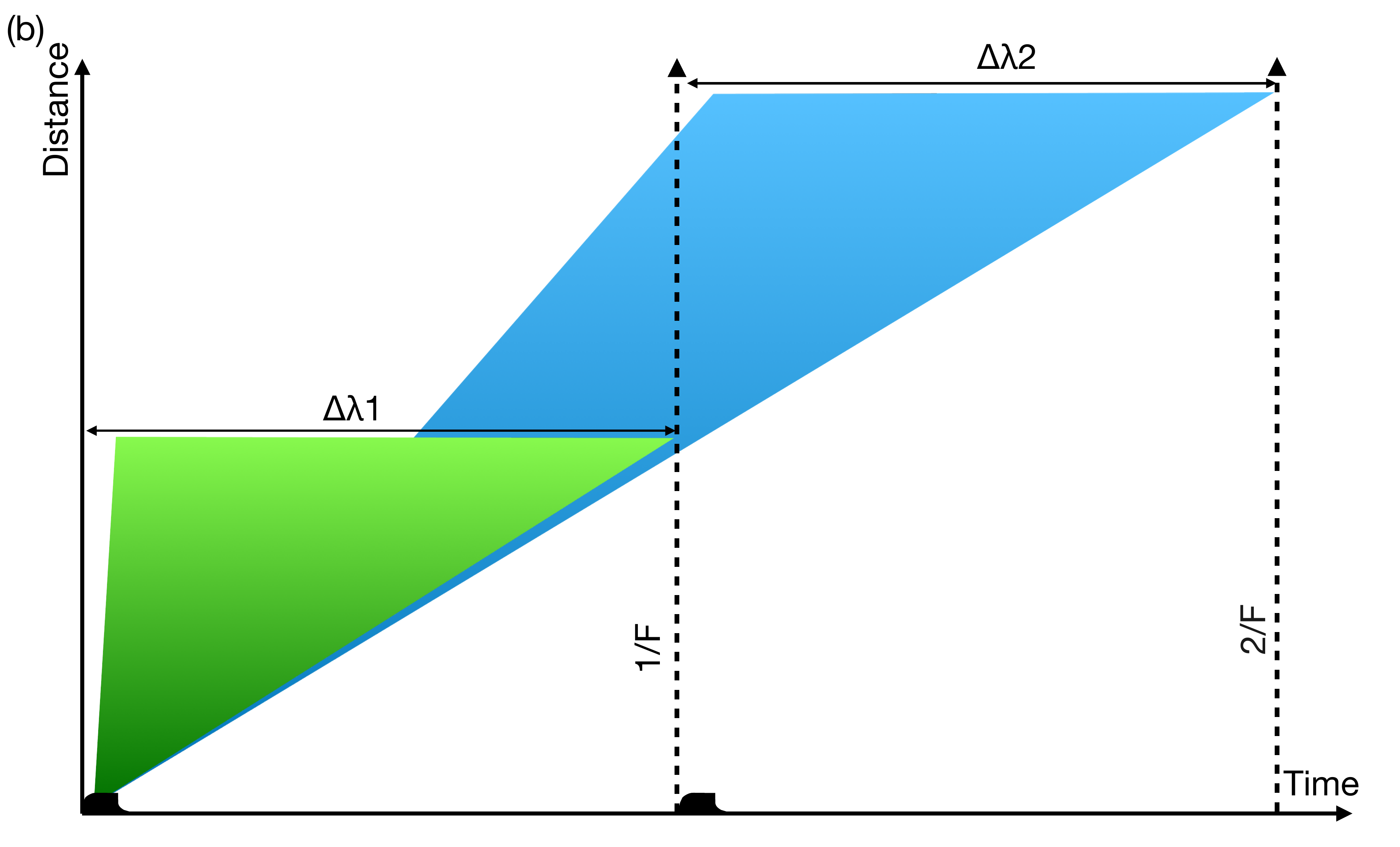}
	\caption{(a) Time-distance diagram showing the relevant parameters for time of flight chopper spectroscopy.(b) The relevant bandwidth accessible on a spectrometer at a spallation source instrument as a function of distance from moderator and period of the spallation source. In the case of cold neutrons, a short instrument will often measure in the first time frame while on a longer instrument the second or third time frame will be used.}
	\label{TimeDistance_Uncertainties}
\end{figure}

A time of flight chopper spectrometer probes the dynamic energy scale of a sample using the time of flight of nearly monochromatic neutrons interacting with a sample and scattered into the detector. The energy resolution of the instrument will determine the lowest energy transfers that can be probed. The energy resolution, $\Delta E$, is determined by the wavelength uncertainty, $\Delta \lambda$, which in turn is determined by the time of flight uncertainty, $\Delta t$, derived from:
\begin{equation}
 E[meV] = \frac{81.81}{\lambda^2[\angstrom^2]}, \lambda[m] = \frac{ht[s]}{m_n[kg] L_{SD}[m]}, 
\label{Eqn1}
\end{equation}
with $L_{SD}$ the sample to detector distance, providing the relationship 
\begin{equation}
t[\mu s] = 252.79\lambda[\angstrom]L_{SD}[m]. 
\label{Eqn2}
\end{equation}

As such the energy resolution at the elastic line is defined by:
\begin{equation}
\frac{\Delta E}{E_{i}} = 2\frac{\Delta \lambda}{\lambda} = \frac{\sqrt E_{i}[meV] \Delta t_{det}[\mu s]}{1142 L_{SD}[m]},
\end{equation}
considering $\Delta\lambda$ extracted from Equation \ref{Eqn2} and $\lambda$ derived from Equation  \ref{Eqn1}. 

The time of flight uncertainty on the detector, $\Delta t_{det}$, is derived from the primary and secondary spectrometer uncertainties, see Figure \ref{TimeDistance_Uncertainties} \cite{Lechner1990}. The primary spectrometer time uncertainties include $\Delta t_{1}$, the uncertainty on the detector due to the pulse shaping (PS) chopper, and $\Delta t_{2}$, the uncertainty on the detector due to the monochromating (M) chopper. A third uncertainty, $\Delta t_{3}$, is related to the the uncertainty in the length of the
secondary flight path caused by finite sample size and detector thickness. The uncertainties are independent and can thus be added in quadrature:
\begin{equation}
 \Delta t_{det} =  \sqrt{\Delta t_{1}^2 + \Delta t_{2}^2 +\Delta t_{3}^2}, 
\end{equation}

$\Delta t_{1}$ and $\Delta t_{2}$ are related to the opening times of the PS and M choppers, respectively, through geometry to the instrument distances: L0, moderator to monochromatic chopper distance, L1, pulse shaping chopper to monochromatic chopper distance, and L2, monochromatic chopper to detector, see Figure \ref{TimeDistance_Uncertainties}(a).  
\begin{equation}
 \Delta t_{1} =  t_{1}\frac{L2}{L1},  \Delta t_{2}  = t2\frac{L2+L0}{L0}. 
\end{equation} 
The flux on the sample,  $\Phi$, for a particular energy resolution at a particular incident wavelength, $\lambda$, can then be calculated according to:
\begin{equation}
\Phi = B \Delta\Omega \Delta\lambda R P,
\end{equation}
with B = source brilliance, $\Delta\Omega$ the divergence accepted in the horizontal and vertical directions and R is the duty cycle, the fraction of time that the beam is on the sample and P is the transport of the guide. R, for a single pulse, is given by t2/ESS Period (71 ms) .  


The European Spallation Source is a long pulsed source with a frequency F = 14 Hz repetition rate, see Figure \ref{TimeDistance_Uncertainties}(b). The wavelength band, $\Delta \lambda$, extracted on a spallation source instrument is inversely proportional to the instrument length according to Equation \ref{Eqn2}. 
Utilization of the available source brilliance in the wavelength band is optimised through RRM. RRM is a common method on many short pulsed spallation source chopper spectrometers, with typically short moderator to detector distance, 20 - 30 m. On these instruments the use of RRM results in 4 - 5 incident monochromatic wavelengths with a broad variation in wavelengths, typically from 2 $<$ $\lambda$ $<$ 10 \angstrom{}, across a single timeframe, $\Delta \lambda_1$ in Figure \ref{TimeDistance_Uncertainties}(b).  The moderator to detector distance on CSPEC is 163.5 m thereby providing a short 1.72 \angstrom{} wavelength band, $\Delta \lambda_2$ in Figure \ref{TimeDistance_Uncertainties}(b), with the possibility to closely match incident wavelengths across the 71 ms (1/14 Hz) period of the ESS. It is easy to see that closely matched wavelengths make it possible, in certain circumstances determined by energy and wavevector transfer resolution requirements, to accumulate pulses and thereby significantly increase flux on sample. It should be noted, Figure \ref{TimeDistance_Uncertainties}(b), that a short instrument will measure cold neutrons in the same time frame as the spallation pulse while on a long instrument the measuring period is a later timeframe. The use of RRM on CSPEC is described in greater detail later in the text.

\section{Quasielastic scattering}

Quasielastic neutron scattering is due to processes occurring with a distribution of energies, such as stochastic rotational motions, reorientations and translational diffusion, on the pico- to microsecond time scales \cite{QENS_Hempelmann}. The aforementioned motions are typically found in hydrogenous materials, soft-matter, biological systems and glassy compounds. Typically, these materials have short range spatial correlations and as such the Q-resolution can be degraded to improve flux, or focus the neutron beam on a smaller sample with a resultant degradation of divergence and thus Q-profiles. In contrast, it is advantageous to have a freely tuneable energy resolution to adjust the experimental time window to the time scales of the motions of interest. The data that one extracts in a quasielastic experiment around zero energy transfer need to be deconvoluted from the experimental shape of the elastic line, the resolution function, and any deviation from a triangular or Gaussian energy resolution function makes it difficult to extract the information required. It is imperative that the signal to noise is optimised and the background is minimized and well understood.  The CSPEC demand for a signal to noise of 10$^{5}$  at 5 \angstrom{} comes from the need to observe weak broad signals, in Q and E, that are relevant in condensed matter. These signals often lie within the background noise and are thus difficult to extract. A high signal to noise will provide the required access. 

 The neutron beam spot size for quasielastic scattering of non-single crystal materials must be able to vary from 4 x 2 cm$^{2}$, thereby optimizing the scattering intensities of dilute but abundant samples, down to 1 x 1 cm$^{2}$ for compounds with small sample volumes. RRM will allow the cumulative use of adjacent incident pulses for incident energy $\lambda$ $>$  6 $\angstrom$, with small variations in energy and momentum transfer, to increase flux. 
 
The comments made in the previous section remain true for quasielastic scattering from a single crystal except that the $\textbf{Q}$-resolution, which is closely matched to the energy transfer, must remain adequate to determine the lifetimes and spatial correlations across the entire S($\textbf{Q}$, $\omega$). The scientific interest in quasielastic scattering of single crystals is rooted firmly in hard condensed matter, typically magnetism or functional materials. Novel materials in these scientific domains are difficult to synthesize and result in small crystals as small as mm$^{3}$. It is therefore important to be able to focus the neutron beam to a small spot size while maintaining a clean divergence profile that can adequately probe S($\textbf{Q}$, $\omega$).

\section{Inelastic neutron scattering}
Inelastic neutron scattering occurs due to processes with discrete energy steps. These include local excitations such as crystal field levels, collective excitations such as phonons and spin waves and higher energy vibrational modes such as stretching modes and librations.

An understanding of crystal field levels can be adequately gained by scattering from a powdered sample  since no variation in energy transfer with wavevector transfer is expected \cite{Cai_Er3Ga5O12, Zub_2018_CEF}. In contrast, collective excitations, with very distinct variations in energy and wavevector transfer, from powdered samples can also provide useful, albeit limited, information by comparing computational modelling methods such as density functional theory or linear spin wave theory with powder averaged S(Q, $\omega$). This is particularly useful for compounds that cannot be synthesised as large single crystals. However, inelastic neutron scattering from single crystal samples provide the most exacting and accurate information via a 4 dimensional S($\textbf{Q}$,$\omega$). This is exemplified by the very active scientific field of research focussed on novel states of matter such as those proposed by the Kiteav spin model on a honeycomb lattice \cite{Kitaev}. These materials harbour Majorana bound states as its excitations and are expected to be used as a component in topological quantum computers \cite{RevModPhys_SpinLiquids_Nayak}. The magnetic excitations of these novel states are low lying, in energy, and highly unusual. Subtle variations in S($\textbf{Q}$, $\omega$) must be determined to quantify and understand the details of the exchange interactions that lead to these novel states \cite{Banerjee_quantumspinliquid}. To gain optimal information from the scattering profiles it will be imperative to have direct access to theoretical models convoluted with the instrumental resolution. A focussing nose, optimising flux for mm to cm sized samples, will increase signal to noise. Signal to noise can be increased further by accumulating 6-10 incident wavelengths via RRM if the energy and $\textbf{Q}$ resolutions permit.

\section{Kinetic measurements}
A combination of the high flux and the multi-energy mode via RRM makes CSPEC most suitable for experiments that investigate time resolved kinetic phenomena, including pump-probe experiments and out of equilibrium behaviour. The high flux via RRM will enable monitoring of transient phenomena with a time resolution of minutes. For cyclic processes, time-resolved stroboscopic measurements will benefit both from the high flux and the multi-energy mode to probe the dynamic response of a compound to an external trigger with a time resolution in the order of milliseconds \cite{Pieper_TimeDependent}. 

\section{Instrumental details}
The scientific requirements for CSPEC, as outlined in the previous sections, leads to the following specifications: 
\begin{itemize}
\item	CSPEC shall extract a wavelength range of 2 – 20 \angstrom. 
\item	The CSPEC guide shall extract flux with +/-1$^{o}$ divergence at 3 \angstrom{} and more, if possible for higher wavelengths. 
\item	CSPEC shall measure in repetition rate multiplication configuration for all measurement modes. 
\item	CSPEC shall be capable of energy resolutions down to and better than $\Delta$E/E = 1.5\% for wavelengths greater than 4 \angstrom{}.
\item	CSPEC shall be capable of momentum transfer resolution $\Delta$Q/Q = 2\%. 
\item	CSPEC shall provide a signal to noise of 10$^{5}$ at 5 \angstrom. Signal to noise is defined as the peak of the intensity at the elastic line of a vanadium sample versus background obtained far away at a time of flight when the background level has been reached. 
\item	The chopper cascade shall ensure that, for each impinging pulse on the sample with energy $E_i$, an energy transfer $\hbar\omega$ =  0.2$E_i$ $<$ $E_i$ $<$ $\infty$ can be measured. 
\item	The neutron beam at the sample position shall illuminate a sample area ranging from 4 $\times$ 2 cm$^2$ to 1 $\times$ 1 cm$^2$ (height $\times$ width)
\item	The detectors provide a detectable angular range of 5$^{o}$< 2$\theta$ < 140$^{o}$ with a sample to detector distance of 3.5 m in the azimuthal plane and a vertical detectable range of +/- 26.5$^{o}$. 
\item	CSPEC shall probe magnetic excitations in magnetic fields up to 12 T. 
\item	The design of the instrument will accommodate the sample environment equipment required to match the science case.
\item Sample environment within the CSPEC scope for the wide range of scientific cases studied on CSPEC should be consistent with the demands of signal to noise.
\item	The systems design shall provide the space and flexibility necessary to host and drive future developments, for instance further focusing optics for smaller sample sizes, possibly integrated into the sample environment, further in-situ sample environment such as secondary characterization (RAMAN or NMR), XYZ polarization analysis.
\end{itemize}

\begin{figure}[h!]
	\centering
	\includegraphics[width=0.5\textwidth]{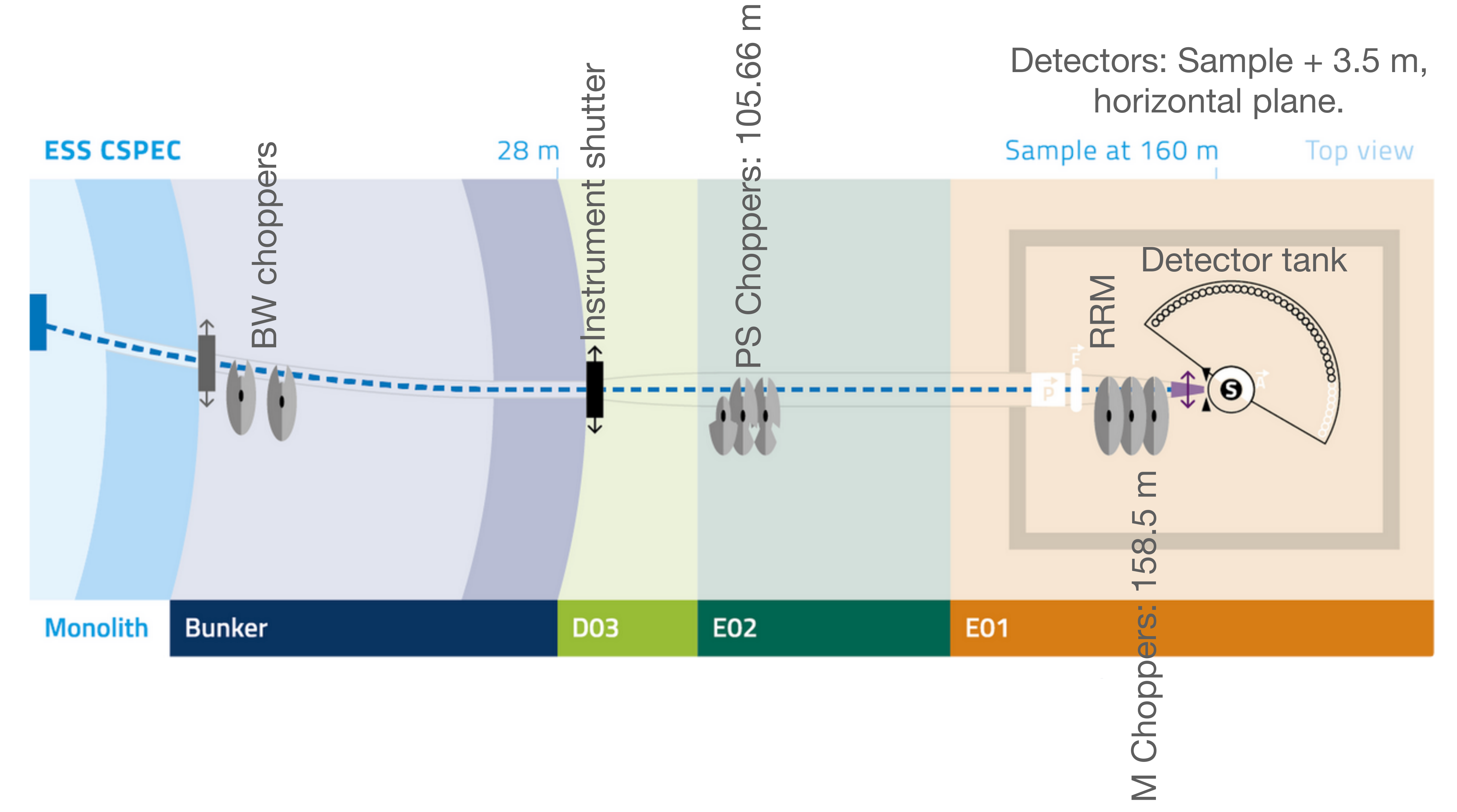}
	\caption{Cartoon overview of the CSPEC instrument showing ESS buildings: Monolith, bunker, D03, E02 and E01 and various CSPEC components and relevant distances. Components in white are upgrade features: polarisation analysis and completion of detector coverage. BW = bandwidth, PS = pulse shaping, M = monochromating choppers \cite{ANDERSEN_ESS_Paper}. Reproduced with permission from Nuclear Instruments and Methods in Physics Research Section A: Accelerators, Spectrometers, Detectors and Associated Equipment
957, 163402 (2020). Copyright Elsevier.}
\label{CSPEC_Overview}
\end{figure}

A cartoon overview of the resultant instrument is shown in Figure \ref{CSPEC_Overview} showing the optics, chopper cascade, instrument shutter, cave, sample environment, detector tank and detector position. 

\section{Instrument Parameters}

\subsection{Chopper cascade}
\begin{figure}[h!]
	\centering
	\includegraphics[width=0.5\textwidth]{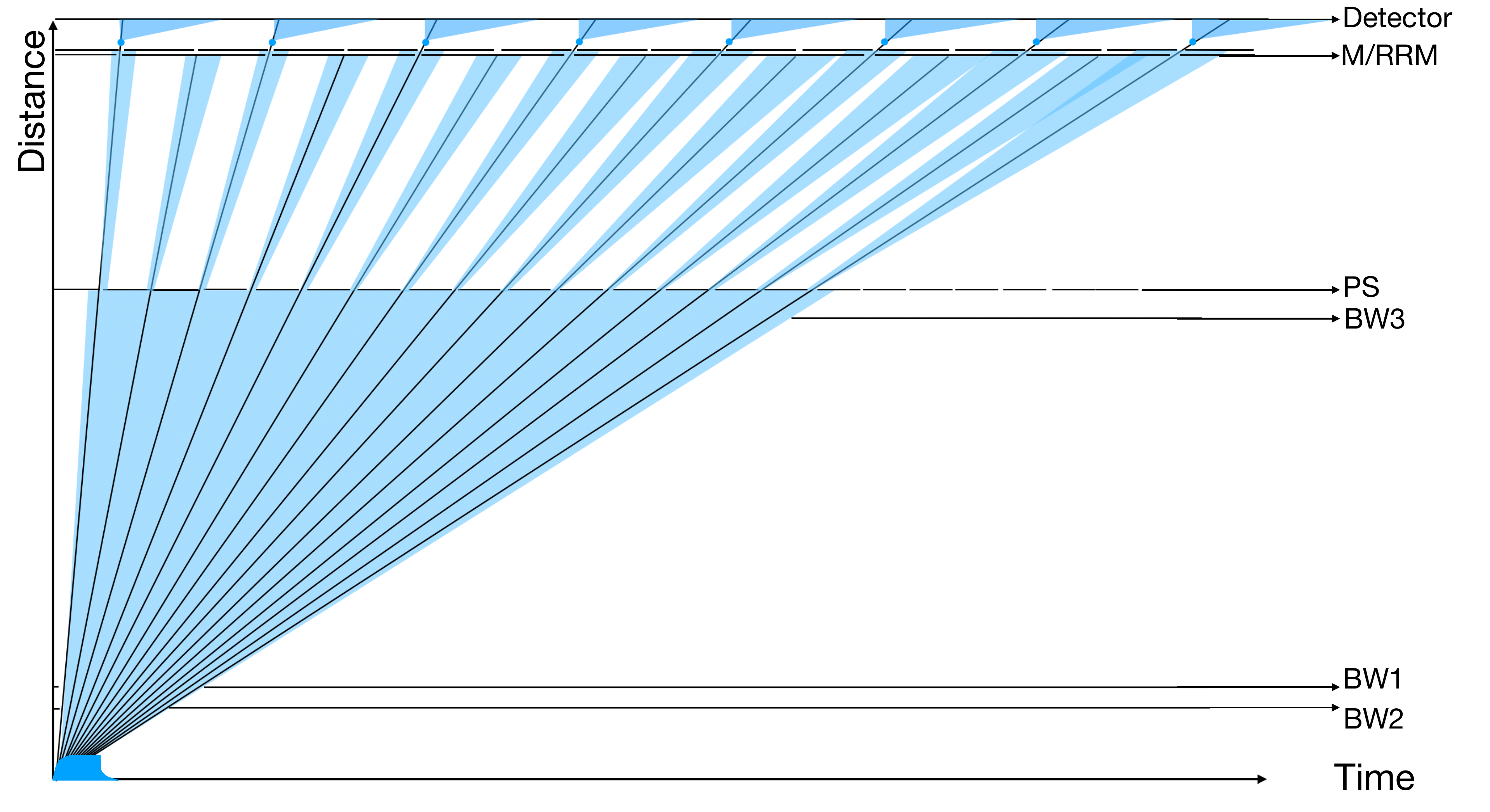}
	\caption{Time-distance diagram defining the chopper cascade on CSPEC.}
	\label{TimeDistance_CSPEC}
\end{figure}

The chopper cascade ensures clean incident monochromatic pulses at the sample position with incident energies E$_i$ and an energy transfer $\hbar\omega$ =  0.2$E_i$ $<$ $E_i$ $<$ $\infty$ measured at the detector position. An energy resolution of $\Delta$E/E = 1.5\% will be achieved for wavelengths equal to or greater than 4 \angstrom. The chopper cascade that will provide these specifications is presented in Figure \ref{TimeDistance_CSPEC} and table \ref{Table_Choppers}. The bandwidth choppers, BW1, BW2 and BW3 have been designed to extract a 1.72 \angstrom{} wavelength bandwidth, around the central wavelength $\lambda_{i}$,  ensuring maximum flux from the time independent part of the ESS pulse structure. The BW chopper openings will be triangular.
The energy resolution chopper cascade, the PS and M choppers, are based on a relative distance and frequency of 2/3 such that the M choppers are positioned at a distance L0 and the distance L1 is x/3 with the PS choppers positioned at 2x/3. To ensure an optimal pulse transmission, through the chopper cascade, the M and PS choppers have one and three transmission windows respectively and rotate at frequencies of F$_{M}$=  2F$_{PS}$,  F$_{M}$ = Frequency of M chopper and F$_{PS}$ = Frequency of PS chopper. The RRM chopper prevents frame overlap between the various incident pulses  and runs at a frequency F$_{M}$/n, n = integer and a multiple of the ESS source frequency.   The chopper openings for the high speed choppers will be rectangular providing a clean scattering profile for the edges, in TOF, of the incident beam.  

\begin{table}
    \begin{tabular}{|l|c|c|c|c|c|}\hline 
       Name  & Pos. (m) & Angle $(^{o})$ & Num & F (Hz) & 	$\varnothing$ (m)   \\ \hline \hline
       BW1 (SB)          & 14.95                    &     40.7 & 1 & 14 & 0.7   \\ \hline
       BW2    (SB)       & 20.47                    &     41.9 & 1 & 14  & 0.7 \\ \hline
       BW3 (SB)          & 104.59                    &     193 & 1 & 14  & 0.7 \\ \hline
       Name  & Pos. (m) & Width $\times$ Height (mm) $(^{o})$ & Num &F (Hz) & 	$\varnothing$ (m)   \\ \hline \hline
       PS (CR)          & 105.67                    &     114.66$\times$ 73.89 & 3 & 168 & 0.7  \\ \hline
    RRM (SB)          & 158.45                    &    22.6 $\times$ 49.0 & 1 & 336 & 0.7  \\ \hline
     M (CR)         & 158.5                    &     22.58 $\times$ 49.0  & 1 & 336 & 0.7  \\ \hline
    \end{tabular}
 \caption{Chopper parameters. Pos = Distance from moderator, Num = number of chopper openings, F = maximum rotation frequency. SB = Single blade, CR = counter rotating blades. $\varnothing$ = blade diameter}
    \label{Table_Choppers}  
\end{table}

\subsection{Guide profile}
Chopper spectroscopy is a flux intensive technique with success based on optimising signal to noise. As such the CSPEC guide must transport cold neutrons with a broad divergence, signal, while limiting high energy fast particles, noise, derived from proton pulse spallation on the target. Liouville's theorem states that an increase in flux per area, beyond the flux extracted from the moderator, is accompanied by an increase in the spread of the beam divergence. The space density of neutron flux extracted from the moderator must be transported to the sample ensuring that the final divergence and flux profile is consistent with the instrument specifications. The divergence profiles of the neutron beam are derived from geometric considerations of the beam extracted from the moderator, the horizontal and vertical guide profiles and the wavelength dependence of the critical angle, $\theta_c$, of neutron reflection of Ni/Ti supermirror guides
\begin{equation}
\theta_c[\deg] = 0.099m\lambda[\angstrom], 
\end{equation}
in which m represents the increase in critical angle of a Ni/Ti supermirror relative to that of a Ni thin film, \cite{Mezei1988}. 
 
High energy particles, known as the prompt pulse,  will be ejected after the proton beam hits the target. The prompt pulse is a time dependent feature that is observed on all spallation instruments and has been shown to extend up to 3 ms beyond the proton pulse on the target with a slow exponential decay consistent with the moderation of fast particles through the facility and instrument structures \cite{Priv_Comm_PersonG_Ehlers}. In the case of the ESS, the proton pulse on target extends 3 ms and therefore there will be a prompt pulse across 6 ms in time. Limiting high energy particle transport through the instrument is achieved by curving the guide away from the line of sight of the target \cite{MEZEI2000318,AmaterasCurvedGuides}. 

	\begin{figure}[h!]
	\centering
	\includegraphics[width=0.35\textwidth]{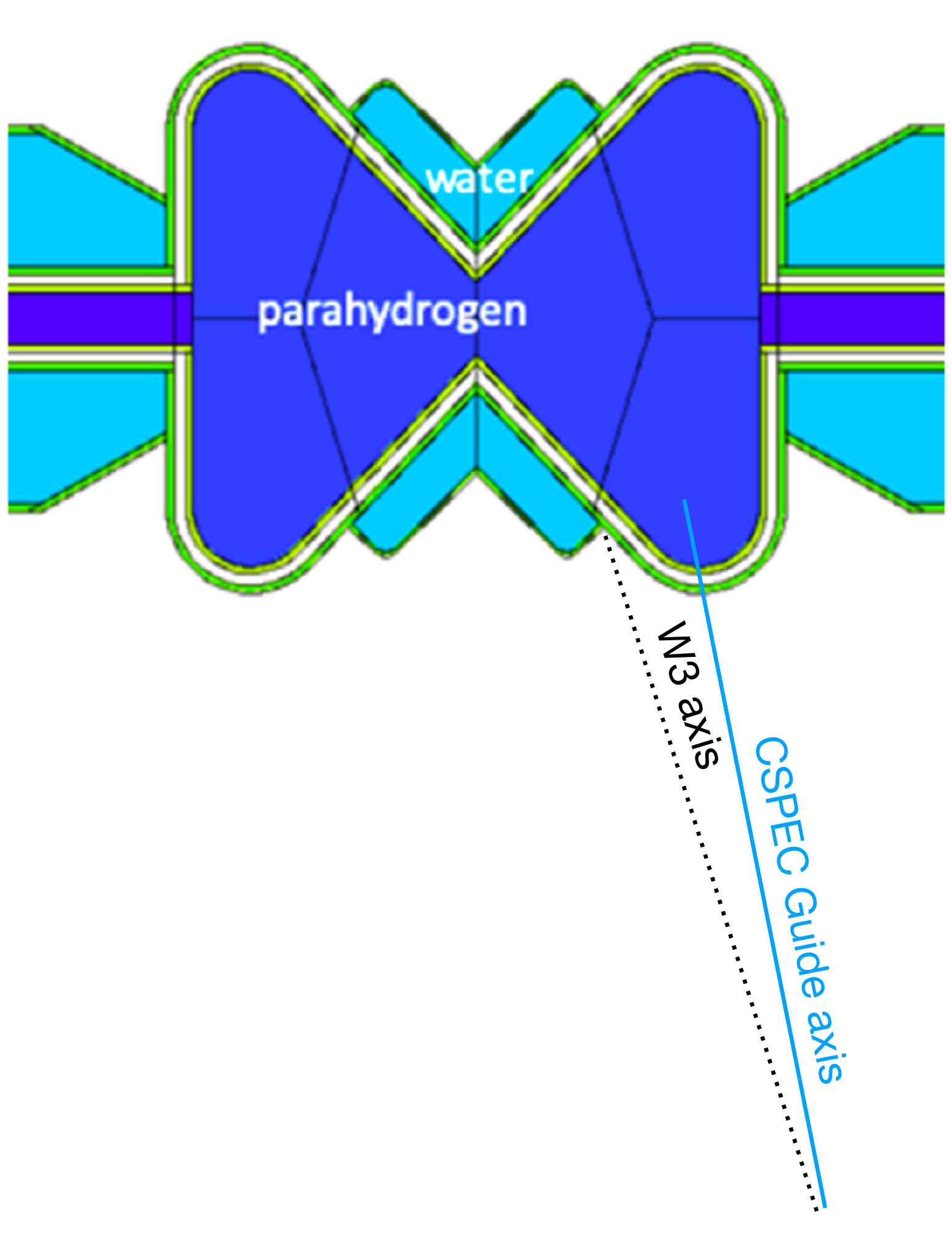}
	\caption{Top view of ESS moderator,\cite{ANDERSEN_ESS_Paper}, annotated with the port axes and instrument axis. Reproduced with permission from Nuclear Instruments and Methods in Physics Research Section A: Accelerators, Spectrometers, Detectors and Associated Equipment 957, 163402 (2020). Copyright Elsevier.}
		\label{Moderator}
	\end{figure}
	
	\begin{figure}[h!]
	\centering
		\includegraphics[width=0.5\textwidth]{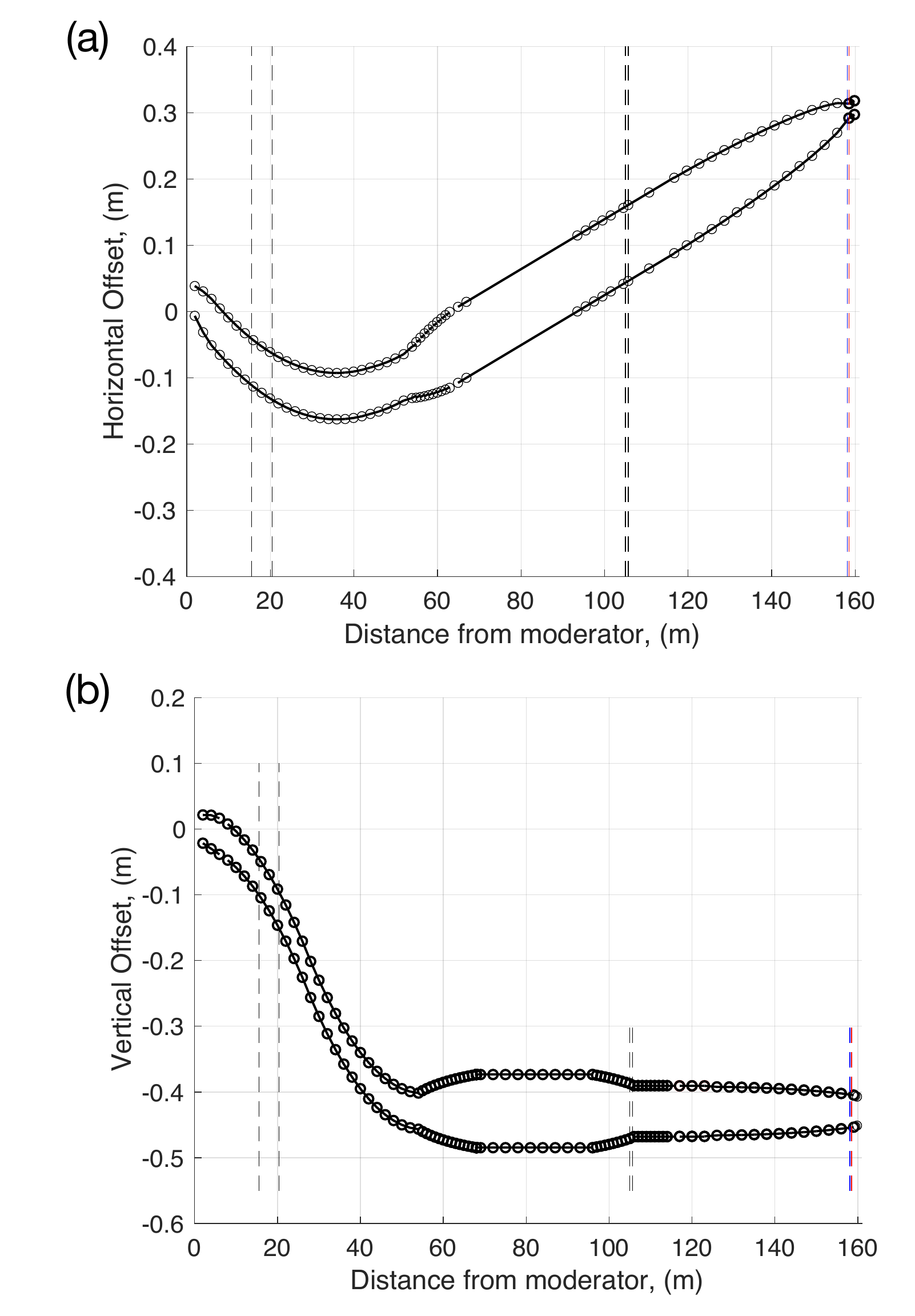}
	\caption{(a) Horizontal and (b) vertical overview of the CSPEC Guide. Dashed lines represent chopper positions.}
	\label{GuideOverview}
\end{figure}

The CSPEC guide starts at 1.9 m from the 3 cm tall moderator using the port axis W3. The W3 port axis is positioned between the thermal and cold moderators. The guide is rotated away from the W3 axis, see Figure \ref{Moderator}, to view only the cold H${_2}$ moderator and thus extract 2 $<$ $\lambda$ $<$  20 \angstrom. The first 2 m of guide sections are funneled from 0.043 m x 0.045 m (height x width)$\rightarrow$ 0.055 x 0.07 m to extract an optimal flux and reduce the divergence profile transmitted beyond the first guide sections. Transporting a reduced divergence will limit the supermirror m-values required for subsequent guide pieces. Limiting the m-value of supermirror guides limits shielding requirements, since the interaction of neutrons with Ni/Ti guides results in high energy gammas, and vastly reduces guide costs. 
The first guide section, 1.9 to 56.4 m, is curved with a radius of curvature (ROC) equal to 1600 m and 4000 m in the vertical and horizontal direction respectively. Vertically, an s-bender guide curves the guide twice out of line of sight (LOS) of the target and moderator. Horizontally, the guide is curved to return the instrument on the W3 instrument axis and to the correct position within the instrument hall, see Figure \ref{GuideOverview}. 
The distance that an instrument will lose line of sight of the moderator is determined by geometric arguments such that
 \begin{equation}
 LOS[m] =  \sqrt{8w[m]R[m]}, 
\end{equation}
 w is the width or height of the curved guide, R is the radius of curvature in the horizontal or vertical direction. The minimum wavelength that is transported by a curved guide is given by $\lambda_c$: 
 \begin{equation}
 \lambda_c{[\angstrom]} =  \frac{575}{m}\sqrt{\frac{2w[m]}{R[m]}}. 
\end{equation} 
Table  \ref{CurvedGuideParams} provides the relevant critical parameters for the CSPEC curved guide. 
\begin{table}[htp]
    \centering
    \vline 
\begin{tabular}{c|c|c|c|c|c|c|c}
\hline
\hline
     Vertical  &     Width (m) & ROC (m) & LOS (m) & $\lambda_c$ (\angstrom) & m-range \\ \hline
        & 0.07   & 1600 & 29.93  & 1.36 & 4 - 3  \\ \hline
        Horizontal  &     Width (m) & ROC (m) & LOS (m) & $\lambda_c$ (Å) & m-range\\ \hline
        & 0.055   & 4000 & 41.95  & 0.972 & 3.5 - 2.5 \\ \hline
        \hline
    \end{tabular}
    \caption{CSPEC Guide parameters for the curved section.}
    \label{CurvedGuideParams}
\end{table}

Beyond the curved guide section an elliptically opening funnel, opening from 0.055 x 0.07 m  (height x width) to  0.111 x 0.115 m (height x width), extends across 14 m and reduces the divergence profile of the beam for the next stretch of guide such that m = 2 supermirrors can be employed. The following 24 m section is a straight section before an elliptical funnel in the vertical direction reduces the height of the beam to 0.074 m at the PS chopper position with m-values slightly increasing to m = 2.5. The final guide section, from the PS to M chopper positions, is again elliptically focussed to achieve the required height and width, 0.049  x 0.023 m  (height x width), to achieve the specified time resolution. M-values increase to m = 3.5 at the chopper position. The final guide section beyond the M-chopper is an exchangeable guide piece that allows both a beam area at the sample position of 4 x 2 cm$^{2}$ and a 1 x 1 cm$^{2}$ (height x width), via an unfocussed tapered and a focussing elliptical nose. The focussing nose provides an increase of 2.5 in flux/cm$^{2}$  across 2 $< \lambda < $ 10 \angstrom{} and a further background reduction achieved by focussing the beam onto a smaller sample area.  

\begin{figure}
    \centering
    \includegraphics[width=0.5\textwidth]{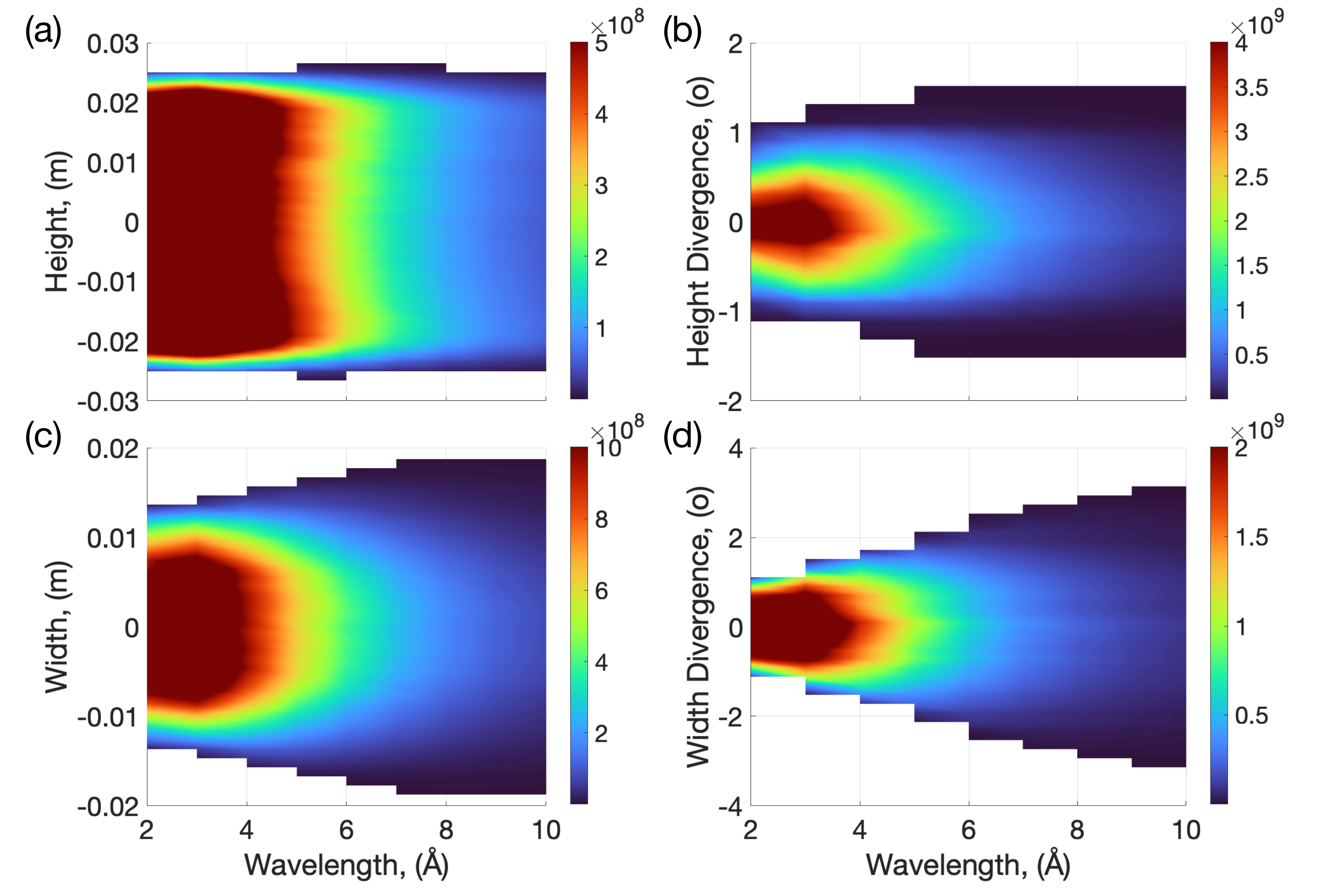}
    \caption{Wavelength dependence of the flux and divergence at the sample position for the unfocussed guide nose showing (a) beam height (b) beam height divergence (c) beam width (d) beam width divergence.}
 \label{Unfocussed_PDF_DIV}
\end{figure}

\begin{figure}
    \centering
    \includegraphics[width=0.5\textwidth]{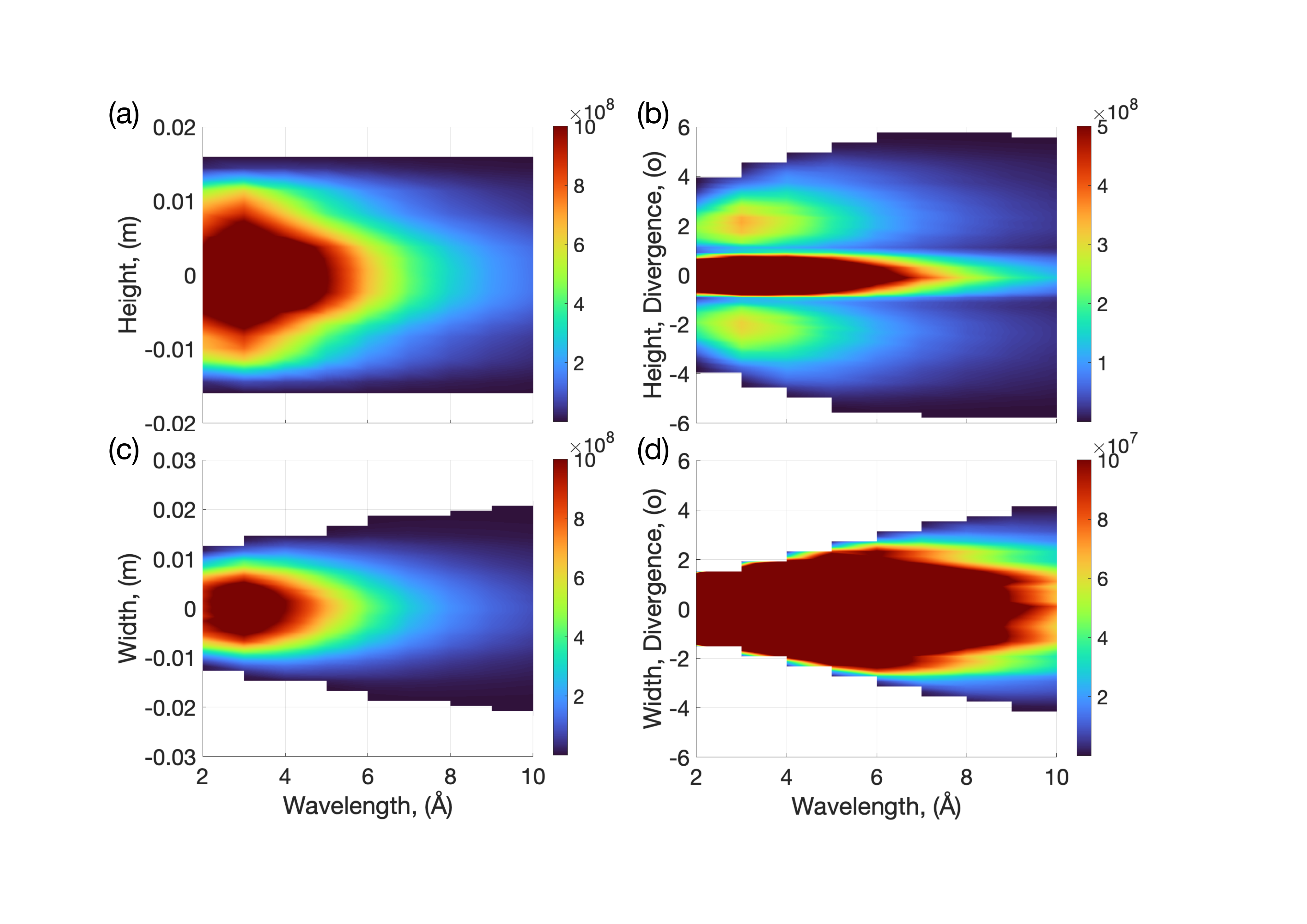}
    \caption{Wavelength dependence of the flux and divergence at the sample position for the focussed guide nose showing (a) beam height (b) beam height divergence (c) beam width (d)beam width divergence.}
  \label{Focussed_PDF_DIV}
\end{figure}

The wavelength dependence of the beam flux and divergence profiles for the unfocussed guide and focussed guide are shown in Figures \ref{Unfocussed_PDF_DIV} and \ref{Focussed_PDF_DIV}, derived from McStas ray tracing simulations \cite{McStas}. The beam profiles for the unfocussed beam are uniform for each wavelength with less than 10 \% variation across the required beam area, 4 x 2 cm$^{2}$. The unfocussed guide provides continuous divergence profiles in both the horizontal and vertical directions. Horizontally, the beam divergence extends to $\pm$ 1$^{o}$ for $\lambda$ $<$ 4 \angstrom{} and  $\pm$ 1.5$^{o}$ for $\lambda$ $>$ 4 \angstrom{}. Vertically, the wavelength dependence of the beam divergence increases linearly with $\pm$ 1$^{o}$ for lambda $=$ 3 \angstrom{} and  $\pm$ 3$^{o}$ for $\lambda$ $=$ 10 \angstrom{}. 

The beam profiles for the focussed beam spot are uniform for each wavelength with less than 10 \% variation across 1 x 0.7 cm$^{2}$. The focussed guide provides a continuous divergence profiles in the horizontal directions with a linear wavelength variation from $\pm$ 1.5$^{o}$ for $\lambda$ $<$ 3 \angstrom{} to $\pm$ 4$^{o}$ for $\lambda$ $=$ 10 \angstrom{}. Vertically the beam divergence shows three peaks, derived from the stringent focussing requirements. This profile will be observed in the Bragg peaks of single crystals samples aswell as well defined excitations for this scattering direction. It will therefore be important to align crystals in suitable directions making use of the increased flux when possible. In addition it will be essential to consider the convolution of the data with the instrumental resolution to extract relevant information. 

\subsection{Repetition Rate Multiplication on CSPEC}

\begin{figure}[htp]
    \centering
    \includegraphics[width=0.5\textwidth]{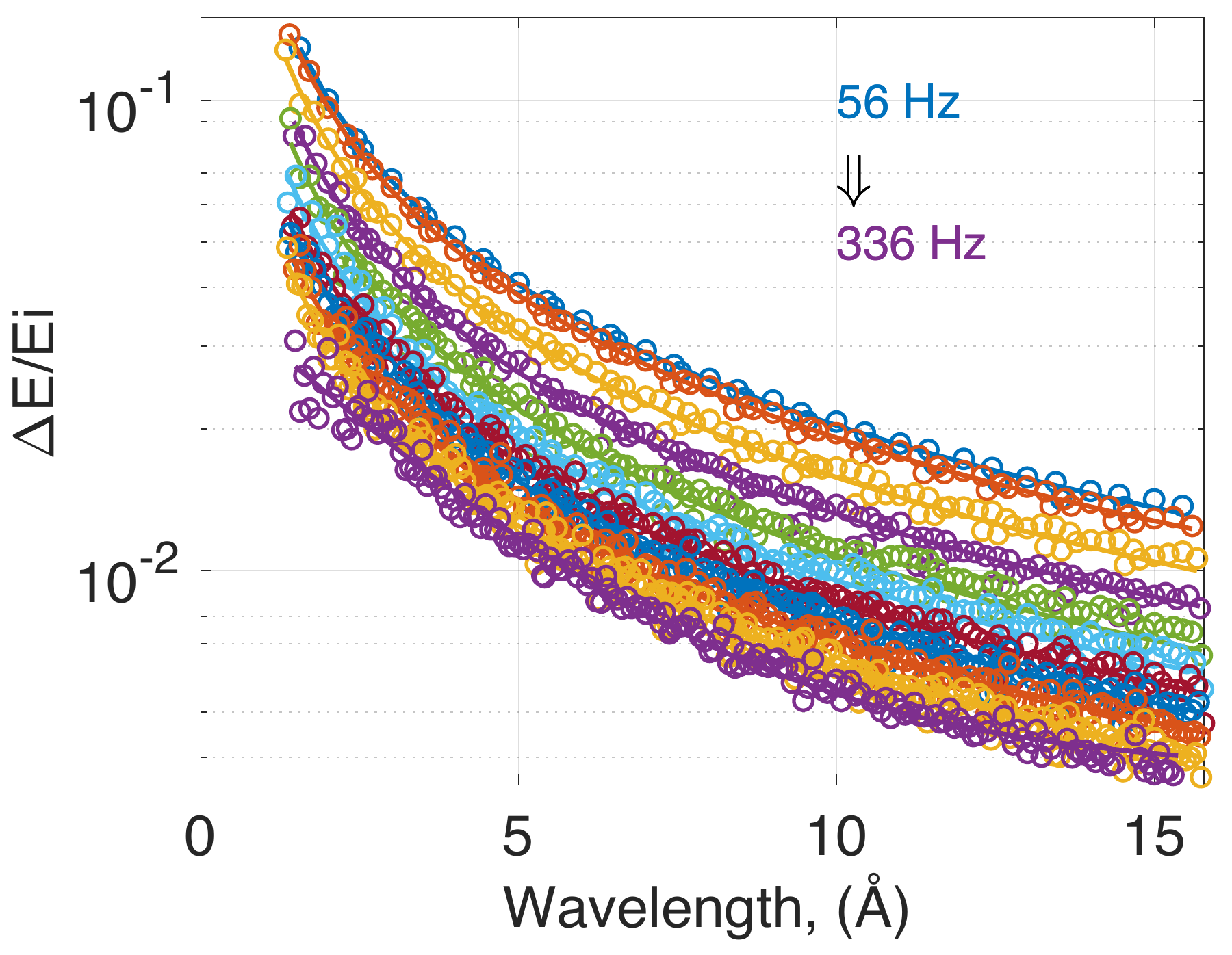}
    \caption{Wavelength dependence of the energy resolution of the elastic line as a function of the rotational frequency of the M chopper. Energy resolution extracted from McStas simulations \cite{McStas}. }
    \label{ERes}
\end{figure}

In this section we will provide an overview of the possibilities afforded by RRM on CSPEC. The energy resolution is defined by the M chopper which is linked to the PS chopper as stated previously: F$_{M}$ = 2 F$_{PS}$.  It is essential that the choppers are phased to the ESS source, 14 Hz, and as such only multiples of the ESS source that are wholly divisible by 2*14 can be employed, i.e. 28:28:336.  The bandwidth of CSPEC is  $\Delta\lambda$ =  1.72 \angstrom{} and incident wavelengths, $\lambda_{i}$, in the range, $\lambda_{i}$-$\Delta \lambda/2$ $<$ $\lambda$ $<$ $\lambda_{i}+\Delta \lambda/2$  will impinge on the sample within one ESS time period. The energy resolution of the elastic line as a function of incident wavelength and M chopper frequency,  determined via McStas simulations \cite{McStas}, can be described by a power law function as shown in Figure \ref{ERes}. It can be seen that $\Delta E/E$ for low $\lambda_i$ varies greatly while $\Delta E/E$ for higher wavelengths are nearly equivalent across $\Delta\lambda$, see Figure \ref{ERes}. 
 
\begin{figure}[htp]
    \centering
    \includegraphics[width=0.5\textwidth]{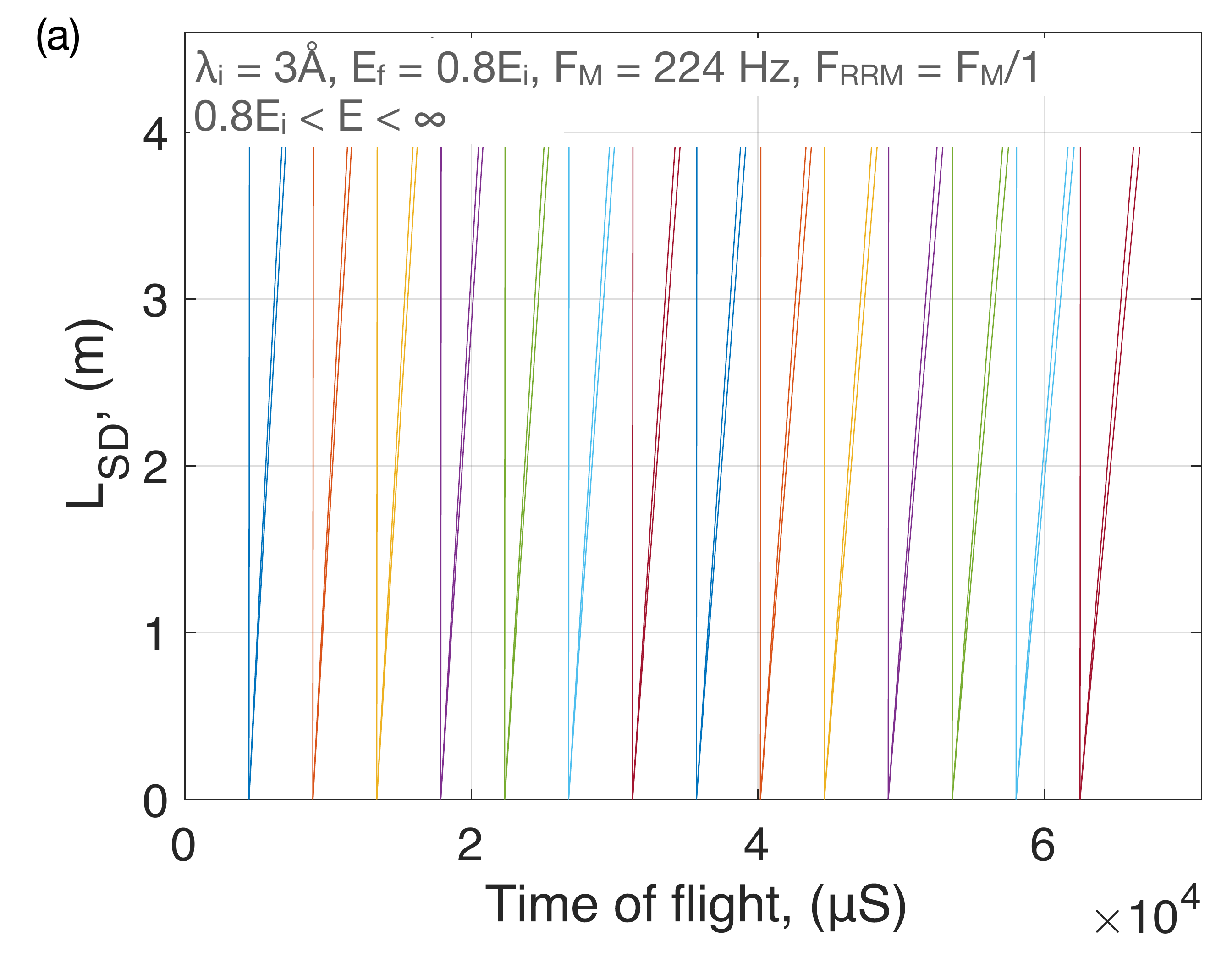}
    \includegraphics[width=0.5\textwidth]{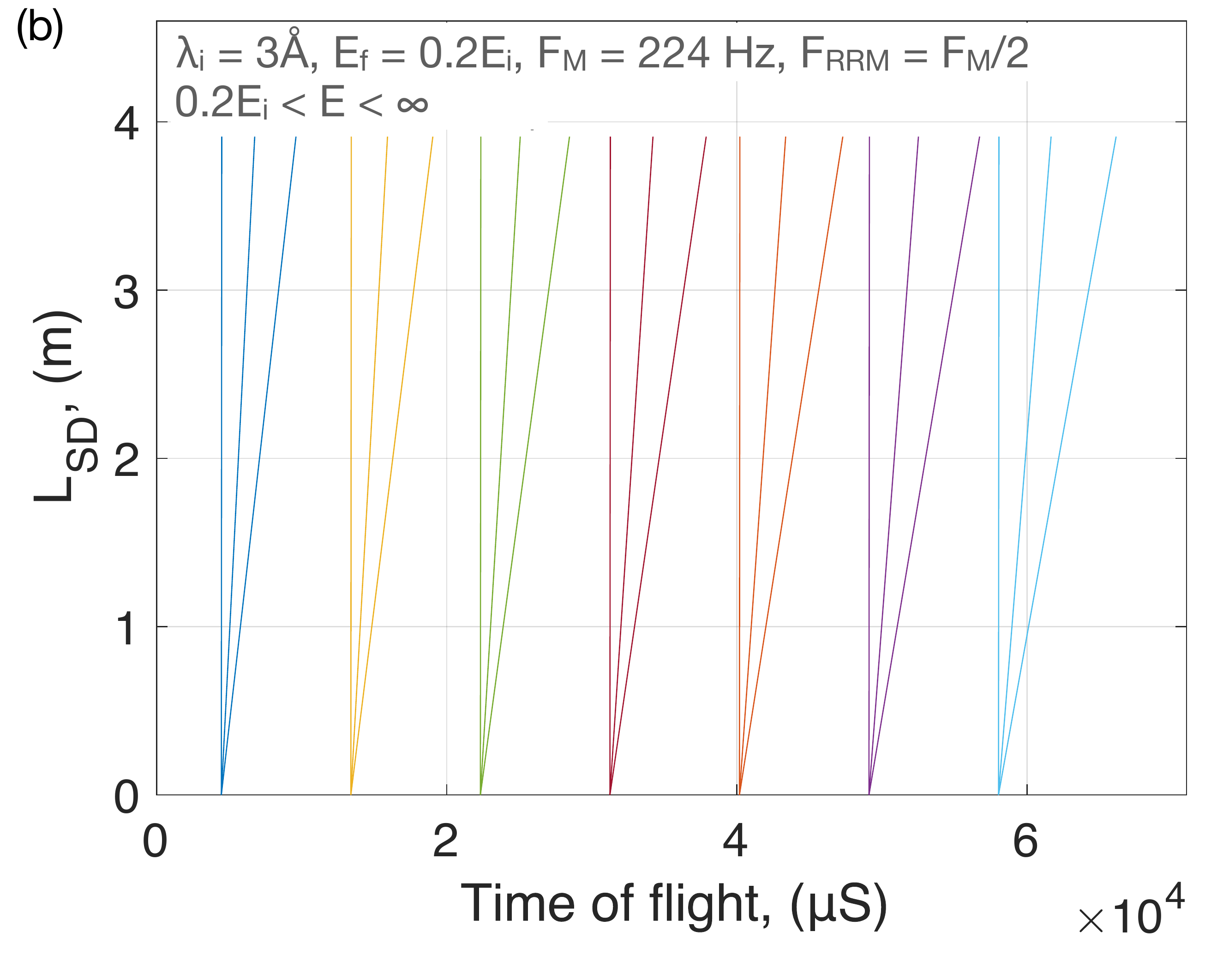}
    \caption{Time distance diagram across the ESS period, 71 ms, for a bandwidth of incident neutrons centered on 3 \angstrom{} scattered from sample (at L$_{SD}$ = 0) to detector. (a) Quasielastic scattering with $E_f$ = 0.8 $E_i$, M chopper rotation = 224 Hz, RRM chopper rotation = 224 Hz.(b) Inelastic scattering with $E_f$ = 0.2 $E_i$, M chopper rotation = 224 Hz, RRM chopper rotation = 112 Hz.}
    \label{IncidentPulses}
\end{figure}

It is up to the experimental team to determine how many pulses they are able to accommodate within their wavevector and energy transfer resolution. Furthermore, the number of pulses that can be accommodated will depend on the energy loss acceptable within the experiment. A quasi-elastic experiment can accommodate an energy loss of $E_f$ = 0.8$E_i$ while an inelastic experiment may need  $E_f$ = 0.2$E_i$, as explained diagrammatically in Figure \ref{IncidentPulses}. Figures \ref{IncidentPulses} shows time of flight diagrams from sample to detector (considering the largest sample - to detector distance = 3.913 m) and the adjacent pulses within an ESS period for (a) $\lambda_i =$ 3 \angstrom{}, with the M and RRM choppers both running at 224 Hz and $E_f$ = 0.8 $E_i$ and (b) $\lambda_i =$ 3 \angstrom, with the M running at 224 Hz and RRM running at 112 Hz to achieve $E_f$ = 0.2 $E_i$.  Frame overlap between adjacent pulses is not an issue in Figure \ref{IncidentPulses}(a) and all possible incident pulses are transmitted, 14 pulses.  In the case of an increased dynamic range, with $E_f$ = 0.2$E_i$, the RRM chopper removes half of the pulses to reduce frame overlap and the number of pulses transmitted is reduced to 7, Figure \ref{IncidentPulses}(b). 

The variation of the energy resolution across the CSPEC bandwidth for 3 \angstrom{} is  = 0.023 $< \Delta E/E < $ 0.039 which may not be scientifically suitable to accumulate. However, for the bandwidth around 8 \angstrom{} the energy resolution varies from 0.01 $< \Delta E/E < $ 0.012 and can be more easily accommodated into a single data set, see Figure \ref{ERes}. 

\subsection{Experimental area}
\begin{figure}[htp]
    \centering
     \includegraphics[width=0.47\textwidth]{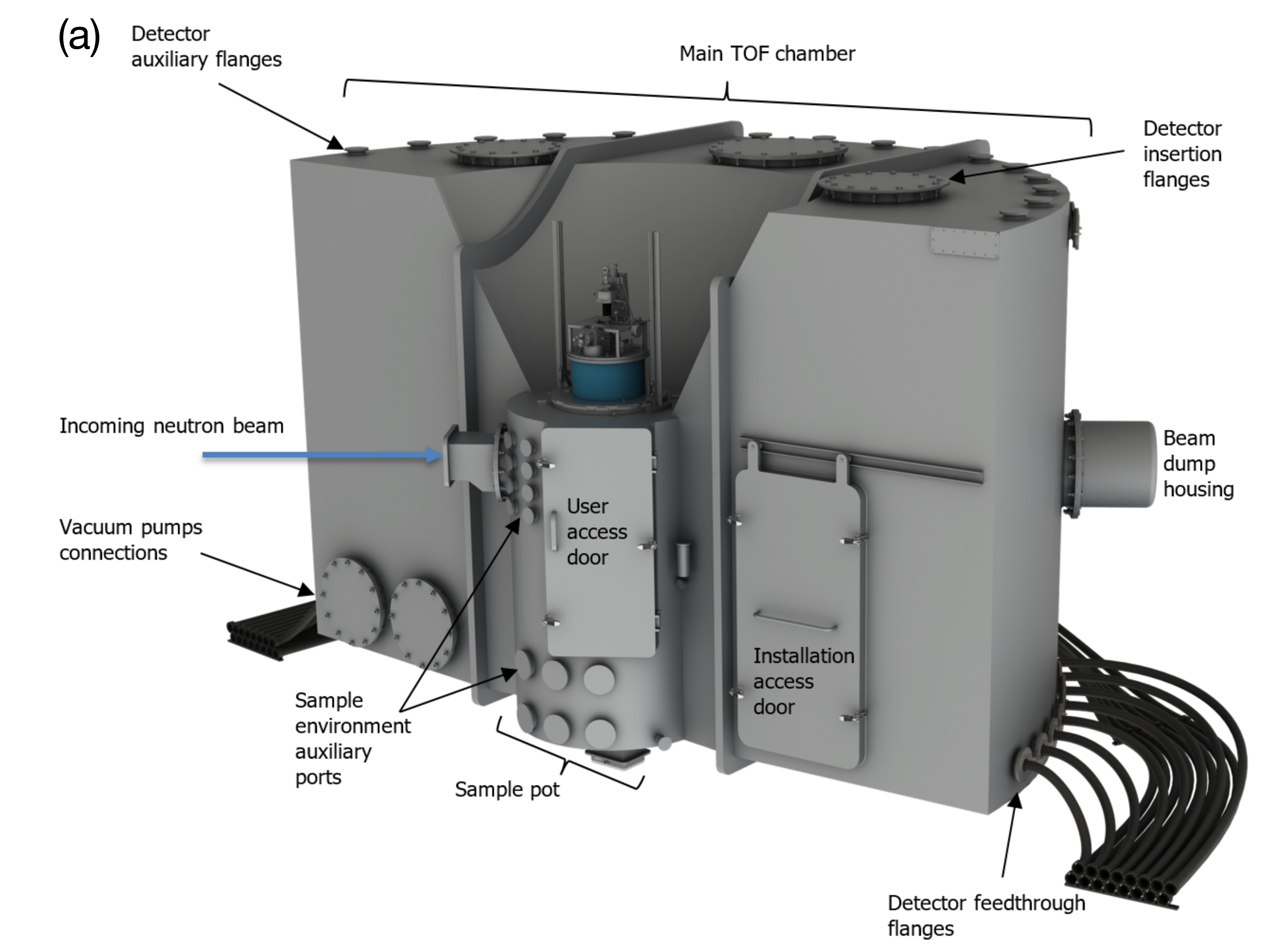}
      \includegraphics[width=0.45\textwidth]{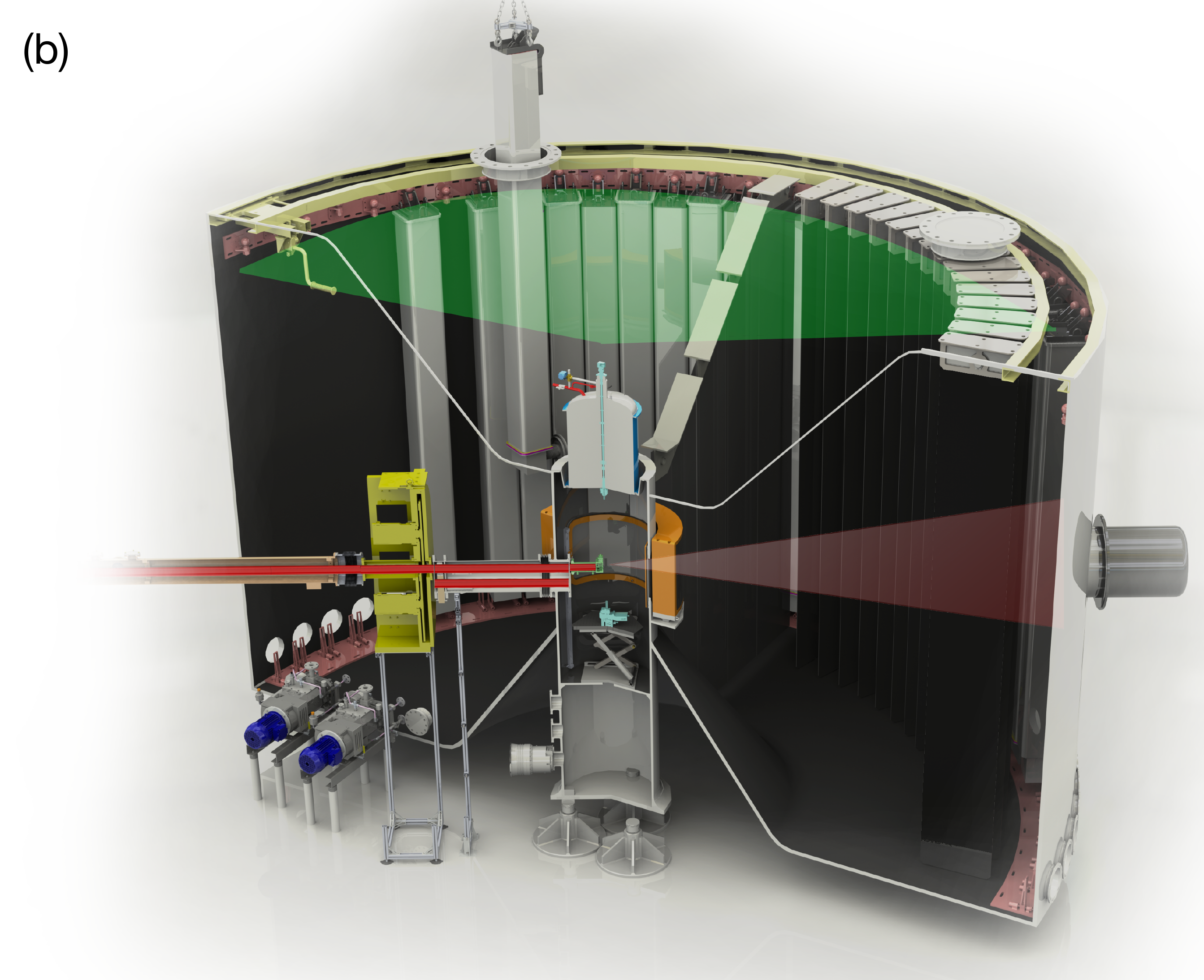}
     \includegraphics[width=0.45\textwidth]{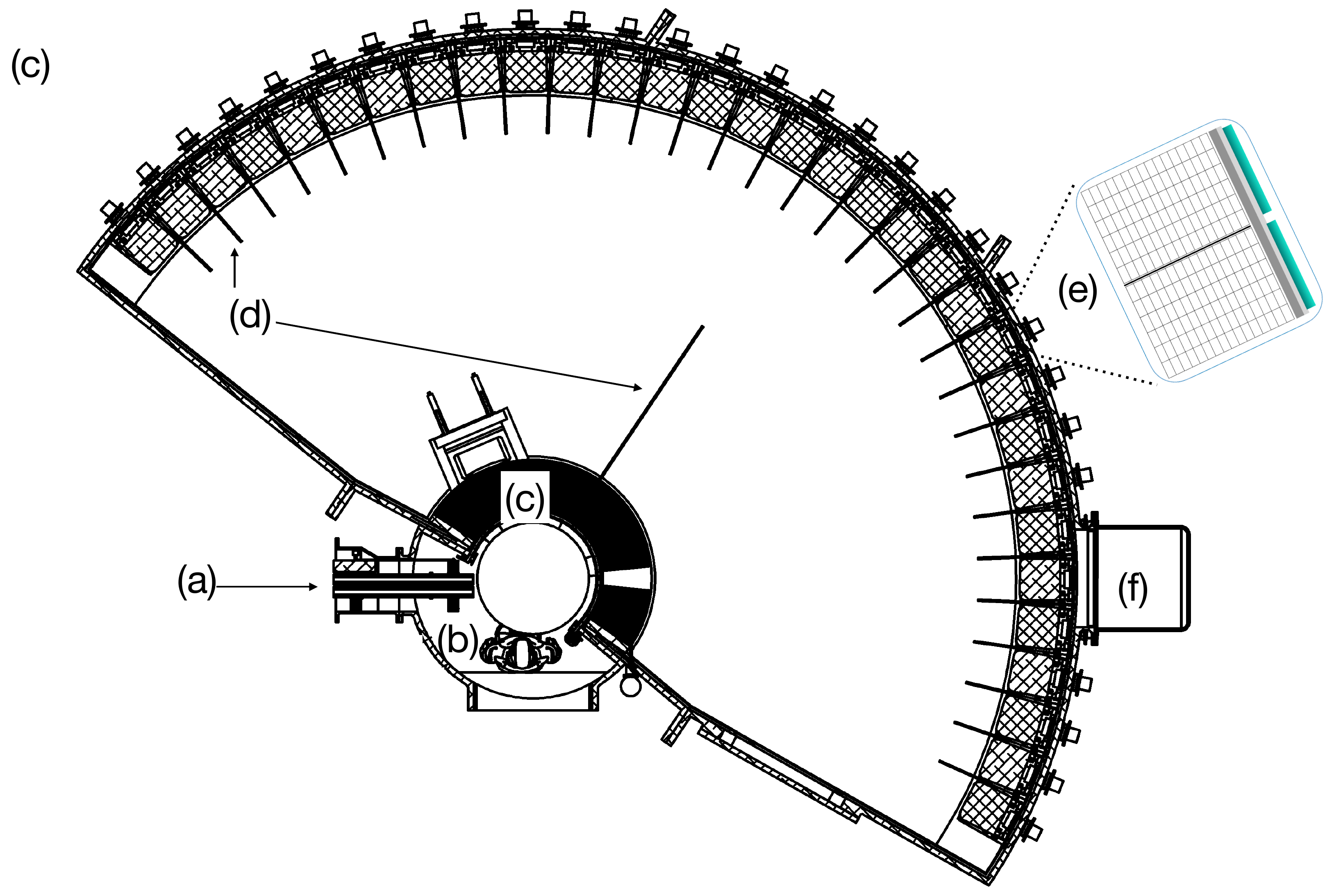}
    \caption{(a) Overview of the detector tank with particular components outlined. (b) Vertical cut through the detector tank. (c) Horizontal cut through the detector tank with an overview of the detector array in the detector tank. (a)incident neutron beam, (b) sample environment pot, with top view of a person, (c) radial collimator (d) detector collimation vanes (e) detector vessel with zoomed image of the two detector columns within a vessel (f) beam dump.The zoomed image shows the 2 detector columns separated by a neutron absorbing vane.}
 \label{fig_DetectorTank_b}
\end{figure}

The CSPEC sample environment area is optimised for the broad range of science that a cold chopper spectrometer addresses. Figure \ref{fig_DetectorTank_b} shows a preliminary engineering drawing of the detector tank inclusive of sample pot. The sample environment area is accessible via a side door and a top flange. Top access is envisaged for conventional cryostats and magnets as sample environment while side access is envisaged for more specialised sample environments. The sample pot diameter is 1 m to provide maximum access. Auxiliary sample environment ports have been added to provide access for further utilities.  Figure ~\ref{fig_DetectorTank_b} shows a cut through the detector tank and reveals, from left to right, the guide exchanger, the sample environment pot, the 4 mm thick Aluminium gate valve, oscillating radial collimator, detector collimation vanes, detectors and beam stop. In this image a detector is being removed for maintenance. The 4 mm thick Aluminium gate valve is designed to easily separate the sample environment pot environment from the evacuated detector area, in the closed position, such that the sample environment pot can provide a different environment to the evacuated detector tank, if an experiment requires it. In normal operation the gate valve will be in the open position and can provide an unimpeded neutron scattering path from sample to detector. 

The detector tank must be consistent with the magnetic field sample environment requirements and the requirements of neutron polarisation, to be implemented shortly after CSPEC comes online. In order to pursue polarisation analysis the magnetic environment that the incident neutrons are affected by must be limited to avoid any depolarisation \cite{Stewart_D7}. The requirement is such that within the sample pot and up to 1.45 m radius from the sample point, the base material, welding seams and all built in components, shall have an ‘as built’ relative magnetic permeability $\mu_{r}$ $<$ 1.01. 

\subsection{Radial collimator}
\begin{figure}[htp]
    \centering
     \includegraphics[width=0.45\textwidth]{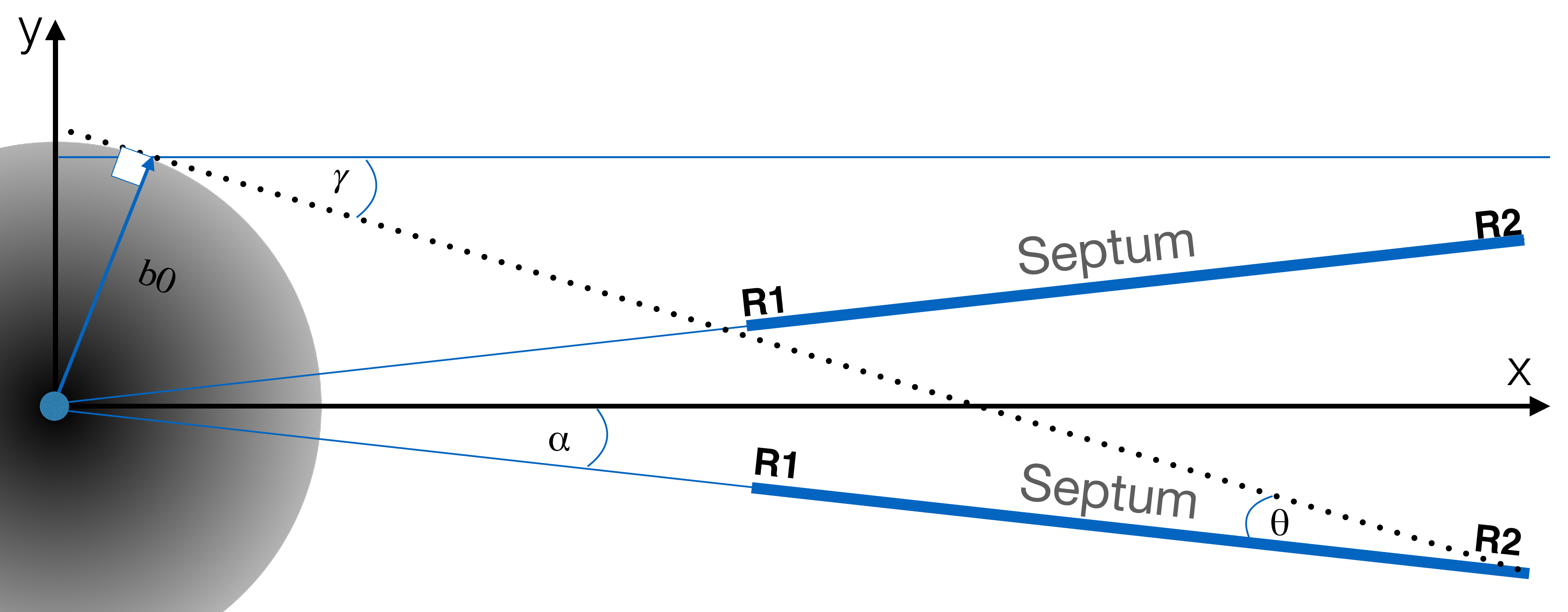}
    \caption{An overview of the relevant parameters for the radial collimator design. The septa start at a distance $R1$ and extend out to a distance $R2$ while separated by an angle of $2\alpha$. Scattering will be absorbed by the septa if the scattering arrives from a distance greater than b0 from the sample and the scattering angle is greater than $\gamma$.}
   \label{fig_RadialCollimator}
\end{figure}

 A radial collimator is a series of septa placed radially,  separated by an angle of $2\alpha$, at a distance $R1$ from the centre of the sample with a length extending to a distance $R2$ from the centre of the sample, see Figure ~\ref{fig_RadialCollimator}. The septa absorb scattering from a spatial and angular region that extends beyond the sample, $>$ $b0$, and for a scattering angle $>$ $\gamma$, respectively. A radial collimator is essential equipment to improve the signal to noise of the experiment. The radial collimator on CSPEC will provide quite tight collimation to limit any secondary scattering effects from sample environments and will be needed to reach the CSPEC specification of a signal to noise of 10$^{5}$.
 
 The CSPEC sample pot diameter must provide sufficient space for a wide range of experiments and this affects the $R1$ parameter of the radial collimator. The minimum radius of the collimator and thus $R1$,  directly beyond the Al gate valve, is 504 mm. Septa thicknesses are not considered in the following considerations.
 
Geometric considerations following Figure \ref{fig_RadialCollimator} results in the following relationships. 
\begin{equation}
    b_0 =  \frac{2 \alpha R1 R2 }{R2-R1}, \\
\end{equation}
\begin{equation}
    \gamma =   \theta + \alpha = \alpha + atan\frac{2R1 sin(\alpha)}{R2-R1}
\end{equation}

Consistency with the CSPEC requirements provides the following radial collimator parameters, see Table \ref{tab:RadCollimator}: 
\begin{table}[htp]
    \centering
    \begin{tabular}{|c|c|c|c|c|}
    \hline
        $R1$ &  $R2$  & 2$\alpha$ & 2$b_0$  & $\gamma$ $^{o}$ \\
        \hline
        \hline
         504 mm & 708 mm & 0.8  $^{o}$ & 48.85 mm &  2.38  $^{o}$\\
         \hline
    \end{tabular}
    \caption{CSPEC radial collimator parameters.}
    \label{tab:RadCollimator}
\end{table}

The radial collimator will oscillate to providing a time averaged scattering profile that is not affected by the septa widths. As such the radial collimator must withstand a 0.1 Hz oscillation with a $\pm$1$^{o}$ oscillation with variable amplitude. The absorbing material must provide a transmission less than 10$^{-5}$. This will be possible by coating thin Mylar foils (micron thickness) with either (99\%B10) B$_{4}$C or Gd$_{3}$O$_{2}$ in an ultra-high vacuum compatible binder.

\subsection{Detector technology}

The detector technology for chopper spectroscopy has historically been based on $^{3}$He absorption technology and has defined the standard for thermal neutron scattering. Indeed, $^{3}$He–based detectors have many of the required characteristics for thermal neutron scattering which includes high uniformity of response, high neutron counting efficiency and low gamma-sensitivity. The technology is very mature and robust. However, the recent $^{3}$He crisis made it necessary to consider novel technologies for neutron scattering research. A novel detector technology for CSPEC is based on thin film converters of Boron-10 Carbide. These new technologies are complex to implement and significant tests have been completed to conform to the CSPEC requirements which include: 
\begin{figure}[htp]
    \centering
 \includegraphics[width=0.5\textwidth]{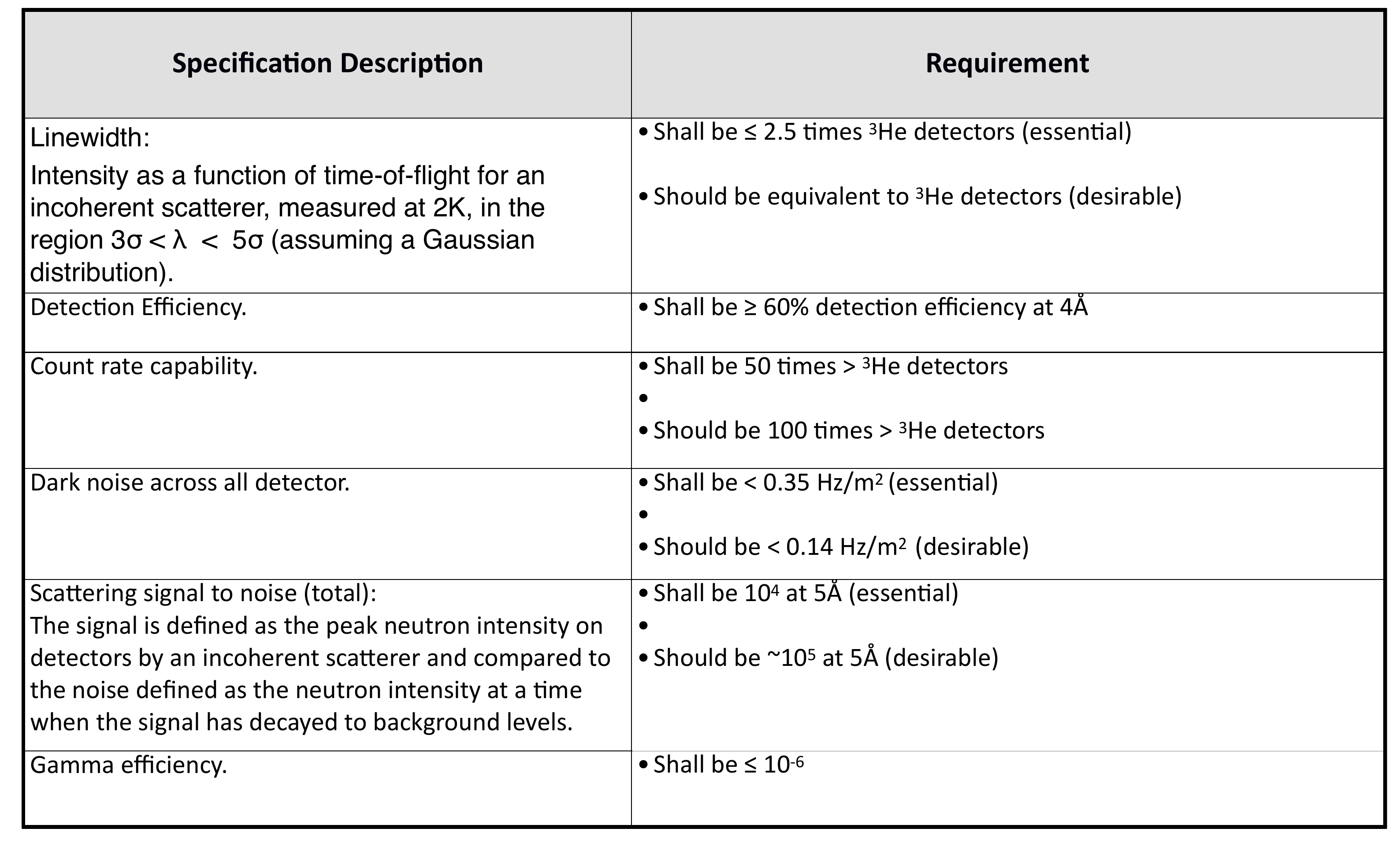}
  \label{tab:DetSpecs}
 \end{figure}

It is beyond the scope of this paper to review the Boron-10 Carbide detector development, these are addressed in a number of different papers, ~\cite{MG_Anastasopoulos_2017} and references within. Geometrically, the detector geometry is distinctly different to a $^{3}$He detector, and is shown in see Figure ~\ref{fig_DetectorTank_b}(c) which shows a top view of the detector tank with the 32 detector vessels placed at the edge of the evacuated detector tank, see Figure ~\ref{fig_DetectorTank_b}(c): (a) the incident neutron beam, (b) the sample environment pot, with top view of a person, (c) the radial collimator, (d) detector collimation vanes, (e) detector vessel with zoomed image of the two detector columns within a vessel and (f) beam dump. The two detector columns in each vessel are separated by a neutron absorbing vane and each detector vessel is separated from the next detector column by a collimation vane. Each detector vessel can be aligned to the sample position to ensure optimal scattering.

A detector column on CSPEC is 3.5 m tall with the detector area subdivided into voxels of 2.5 $\times$ 2.5 $\times$ 1 cm$^{3}$ providing the required wavevector transfer. Uniquely, for a thermal neutron detector optimised for time of flight spectroscopy, the detector has a significant depth, with 16 $\times$ 1 cm deep voxels to increase detection efficiency. The thickness of the Boron layers, coating the voxels, have been optimised with respect to the CSPEC wavelength range.

Each detector vessel, housing 2 detector columns, will provide 6 (wide) $\times$ 16 (deep) $\times$ 140 (high) $\times$ 2 (columns) equalling 26880 voxels per detector vessel and 860 160 detector voxels for the complete detector array, 2.49 Str and 30.6 m$^{2}$. The extraction of S($\textbf{Q}$, $\omega$) from such a large detector array is bringing about a paradigm shift for data analysis. The details will be published in future works.

\subsection{Conclusion}
In conclusion, an overview of the cold chopper spectrometer of the ESS, CSPEC, has been provided. We show the relevant instrument parameters from moderator to detector tank with emphasis on extracting cold neutrons at a long pulsed spallation source. The instrument specifications and relevant solutions to achieve those specifications are outlined. It will be of great interest to provide a direct comparison of the expected performance to true experimental performance once CSPEC is operational in the coming years.  
\subsection{Acknowledgements}
We thank Peter Link, TUM, for the development and manufacture of the CSPEC optics, in addition to fruitful discussions thereof.
CSPEC is a German/ French collaboration supported equally by the German and French in-kind contributions to the ESS. 
\subsection{Data Availability}
The data that support the findings of this study are available from the corresponding author upon reasonable request.

\nocite{*}
\bibliography{aipsamp}

\providecommand{\noopsort}[1]{}\providecommand{\singleletter}[1]{#1}%
\begin{thebibliography}{30}%
\makeatletter
\providecommand \@ifxundefined [1]{%
 \@ifx{#1\undefined}
}%
\providecommand \@ifnum [1]{%
 \ifnum #1\expandafter \@firstoftwo
 \else \expandafter \@secondoftwo
 \fi
}%
\providecommand \@ifx [1]{%
 \ifx #1\expandafter \@firstoftwo
 \else \expandafter \@secondoftwo
 \fi
}%
\providecommand \natexlab [1]{#1}%
\providecommand \enquote  [1]{``#1''}%
\providecommand \bibnamefont  [1]{#1}%
\providecommand \bibfnamefont [1]{#1}%
\providecommand \citenamefont [1]{#1}%
\providecommand \href@noop [0]{\@secondoftwo}%
\providecommand \href [0]{\begingroup \@sanitize@url \@href}%
\providecommand \@href[1]{\@@startlink{#1}\@@href}%
\providecommand \@@href[1]{\endgroup#1\@@endlink}%
\providecommand \@sanitize@url [0]{\catcode `\\12\catcode `\$12\catcode
  `\&12\catcode `\#12\catcode `\^12\catcode `\_12\catcode `\%12\relax}%
\providecommand \@@startlink[1]{}%
\providecommand \@@endlink[0]{}%
\providecommand \url  [0]{\begingroup\@sanitize@url \@url }%
\providecommand \@url [1]{\endgroup\@href {#1}{\urlprefix }}%
\providecommand \urlprefix  [0]{URL }%
\providecommand \Eprint [0]{\href }%
\providecommand \doibase [0]{https://doi.org/}%
\providecommand \selectlanguage [0]{\@gobble}%
\providecommand \bibinfo  [0]{\@secondoftwo}%
\providecommand \bibfield  [0]{\@secondoftwo}%
\providecommand \translation [1]{[#1]}%
\providecommand \BibitemOpen [0]{}%
\providecommand \bibitemStop [0]{}%
\providecommand \bibitemNoStop [0]{.\EOS\space}%
\providecommand \EOS [0]{\spacefactor3000\relax}%
\providecommand \BibitemShut  [1]{\csname bibitem#1\endcsname}%
\let\auto@bib@innerbib\@empty
\bibitem [{\citenamefont {Mu}\ \emph {et~al.}(2019)\citenamefont {Mu},
  \citenamefont {Hermann}, \citenamefont {Gorsse}, \citenamefont {Zhao},
  \citenamefont {Manley}, \citenamefont {Fishman},\ and\ \citenamefont
  {Lindsay}}]{PhysRevMaterials.3.025403_Phonons_Magnons}%
  \BibitemOpen
  \bibfield  {author} {\bibinfo {author} {\bibfnamefont {S.}~\bibnamefont
  {Mu}}, \bibinfo {author} {\bibfnamefont {R.~P.}\ \bibnamefont {Hermann}},
  \bibinfo {author} {\bibfnamefont {S.}~\bibnamefont {Gorsse}}, \bibinfo
  {author} {\bibfnamefont {H.}~\bibnamefont {Zhao}}, \bibinfo {author}
  {\bibfnamefont {M.~E.}\ \bibnamefont {Manley}}, \bibinfo {author}
  {\bibfnamefont {R.~S.}\ \bibnamefont {Fishman}},\ and\ \bibinfo {author}
  {\bibfnamefont {L.}~\bibnamefont {Lindsay}},\ }\bibfield  {title} {\enquote
  {\bibinfo {title} {Phonons, magnons, and lattice thermal transport in
  antiferromagnetic semiconductor mnte},}\ }\href
  {https://doi.org/10.1103/PhysRevMaterials.3.025403} {\bibfield  {journal}
  {\bibinfo  {journal} {Phys. Rev. Materials}\ }\textbf {\bibinfo {volume}
  {3}},\ \bibinfo {pages} {025403} (\bibinfo {year} {2019})}\BibitemShut
  {NoStop}%
\bibitem [{\citenamefont {Cedervall}\ \emph {et~al.}(2019)\citenamefont
  {Cedervall}, \citenamefont {Andersson}, \citenamefont {Delczeg-Czirjak},
  \citenamefont {Iu\ifmmode~\mbox{\c{s}}\else \c{s}\fi{}an}, \citenamefont
  {Pereiro}, \citenamefont {Roy}, \citenamefont {Ericsson}, \citenamefont
  {H\"aggstr\"om}, \citenamefont {Lohstroh}, \citenamefont {Mutka},
  \citenamefont {Sahlberg}, \citenamefont {Nordblad},\ and\ \citenamefont
  {Deen}}]{PhysRevB.99.174437Fe2P}%
  \BibitemOpen
  \bibfield  {author} {\bibinfo {author} {\bibfnamefont {J.}~\bibnamefont
  {Cedervall}}, \bibinfo {author} {\bibfnamefont {M.~S.}\ \bibnamefont
  {Andersson}}, \bibinfo {author} {\bibfnamefont {E.~K.}\ \bibnamefont
  {Delczeg-Czirjak}}, \bibinfo {author} {\bibfnamefont {D.}~\bibnamefont
  {Iu\ifmmode~\mbox{\c{s}}\else \c{s}\fi{}an}}, \bibinfo {author}
  {\bibfnamefont {M.}~\bibnamefont {Pereiro}}, \bibinfo {author} {\bibfnamefont
  {P.}~\bibnamefont {Roy}}, \bibinfo {author} {\bibfnamefont {T.}~\bibnamefont
  {Ericsson}}, \bibinfo {author} {\bibfnamefont {L.}~\bibnamefont
  {H\"aggstr\"om}}, \bibinfo {author} {\bibfnamefont {W.}~\bibnamefont
  {Lohstroh}}, \bibinfo {author} {\bibfnamefont {H.}~\bibnamefont {Mutka}},
  \bibinfo {author} {\bibfnamefont {M.}~\bibnamefont {Sahlberg}}, \bibinfo
  {author} {\bibfnamefont {P.}~\bibnamefont {Nordblad}},\ and\ \bibinfo
  {author} {\bibfnamefont {P.~P.}\ \bibnamefont {Deen}},\ }\bibfield  {title}
  {\enquote {\bibinfo {title} {Magnetocaloric effect in
  ${\mathrm{fe}}_{2}\mathrm{P}$: Magnetic and phonon degrees of freedom},}\
  }\href {https://doi.org/10.1103/PhysRevB.99.174437} {\bibfield  {journal}
  {\bibinfo  {journal} {Phys. Rev. B}\ }\textbf {\bibinfo {volume} {99}},\
  \bibinfo {pages} {174437} (\bibinfo {year} {2019})}\BibitemShut {NoStop}%
\bibitem [{\citenamefont {Gingras~MJ}(2014)}]{Gingras_QuantumFluctuctuations}%
  \BibitemOpen
  \bibfield  {author} {\bibinfo {author} {\bibfnamefont {M.~P.}\ \bibnamefont
  {Gingras~MJ}},\ }\bibfield  {title} {\enquote {\bibinfo {title} {Quantum spin
  ice: a search for gapless quantum spin liquids in pyrochlore magnets.}}\
  }\href@noop {} {\bibfield  {journal} {\bibinfo  {journal} {Rep Prog Phys.}\
  }\textbf {\bibinfo {volume} {77}},\ \bibinfo {pages} {056501} (\bibinfo
  {year} {2014})}\BibitemShut {NoStop}%
\bibitem [{\citenamefont {Smith}\ and\ \citenamefont {{van
  Gunsteren}}(1994)}]{SMITH_Proteins}%
  \BibitemOpen
  \bibfield  {author} {\bibinfo {author} {\bibfnamefont {P.~E.}\ \bibnamefont
  {Smith}}\ and\ \bibinfo {author} {\bibfnamefont {W.~F.}\ \bibnamefont {{van
  Gunsteren}}},\ }\bibfield  {title} {\enquote {\bibinfo {title} {Translational
  and rotational diffusion of proteins},}\ }\href
  {https://doi.org/https://doi.org/10.1006/jmbi.1994.1172} {\bibfield
  {journal} {\bibinfo  {journal} {Journal of Molecular Biology}\ }\textbf
  {\bibinfo {volume} {236}},\ \bibinfo {pages} {629 -- 636} (\bibinfo {year}
  {1994})}\BibitemShut {NoStop}%
\bibitem [{\citenamefont {Hamaneh}, \citenamefont {Zhang},\ and\ \citenamefont
  {Buck}(2011)}]{HAMANEH_Proteins}%
  \BibitemOpen
  \bibfield  {author} {\bibinfo {author} {\bibfnamefont {M.~B.}\ \bibnamefont
  {Hamaneh}}, \bibinfo {author} {\bibfnamefont {L.}~\bibnamefont {Zhang}},\
  and\ \bibinfo {author} {\bibfnamefont {M.}~\bibnamefont {Buck}},\ }\bibfield
  {title} {\enquote {\bibinfo {title} {A direct coupling between global and
  internal motions in a single domain protein? md investigation of extreme
  scenarios},}\ }\href
  {https://doi.org/https://doi.org/10.1016/j.bpj.2011.05.041} {\bibfield
  {journal} {\bibinfo  {journal} {Biophysical Journal}\ }\textbf {\bibinfo
  {volume} {101}},\ \bibinfo {pages} {196 -- 204} (\bibinfo {year}
  {2011})}\BibitemShut {NoStop}%
\bibitem [{\citenamefont {Boothroyd}(2020)}]{Boothroyd2020}%
  \BibitemOpen
  \bibfield  {author} {\bibinfo {author} {\bibfnamefont {A.}~\bibnamefont
  {Boothroyd}},\ }\href@noop {} {\emph {\bibinfo {title} {Principles of neutron
  scattering from condensed matter}}}\ (\bibinfo  {publisher} {Oxford
  University Press},\ \bibinfo {year} {2020})\BibitemShut {NoStop}%
\bibitem [{\citenamefont {{Ollivier, J.}}\ and\ \citenamefont {{Zanotti,
  J.-M.}}(2010)}]{OllivierIN5}%
  \BibitemOpen
  \bibfield  {author} {\bibinfo {author} {\bibnamefont {{Ollivier, J.}}}\ and\
  \bibinfo {author} {\bibnamefont {{Zanotti, J.-M.}}},\ }\bibfield  {title}
  {\enquote {\bibinfo {title} {Diffusion in\'elastique de neutrons par temps de
  vol},}\ }\href {https://doi.org/10.1051/sfn/2010006} {\bibfield  {journal}
  {\bibinfo  {journal} {JDN}\ }\textbf {\bibinfo {volume} {10}},\ \bibinfo
  {pages} {379--423} (\bibinfo {year} {2010})}\BibitemShut {NoStop}%
\bibitem [{\citenamefont {Lechner}(1990)}]{Lechner1990}%
  \BibitemOpen
  \bibfield  {author} {\bibinfo {author} {\bibfnamefont {R.~E.}\ \bibnamefont
  {Lechner}},\ }\bibfield  {title} {\enquote {\bibinfo {title} {Optimization of
  a multi-disk chopper spectrometer for cold neutron scattering experiments},}\
  }\href@noop {} {\bibfield  {journal} {\bibinfo  {journal} {ICANS-XI
  International Collaboration on Advanced Neutron Sources}\ ,\ \bibinfo {pages}
  {717}} (\bibinfo {year} {1990})}\BibitemShut {NoStop}%
\bibitem [{\citenamefont {Ollivier}\ and\ \citenamefont
  {Mutka}(2011)}]{IN5_ILL}%
  \BibitemOpen
  \bibfield  {author} {\bibinfo {author} {\bibfnamefont {J.}~\bibnamefont
  {Ollivier}}\ and\ \bibinfo {author} {\bibfnamefont {H.}~\bibnamefont
  {Mutka}},\ }\bibfield  {title} {\enquote {\bibinfo {title} {In5 cold neutron
  time-of-flight spectrometer, prepared to tackle single crystal
  spectroscopy},}\ }\href {https://doi.org/10.1143/JPSJS.80SB.SB003} {\bibfield
   {journal} {\bibinfo  {journal} {Journal of the Physical Society of Japan}\
  }\textbf {\bibinfo {volume} {80}},\ \bibinfo {pages} {SB003} (\bibinfo {year}
  {2011})}\BibitemShut {NoStop}%
\bibitem [{\citenamefont {Lohstroh}\ and\ \citenamefont
  {Evenson}(2015)}]{TOFTOF}%
  \BibitemOpen
  \bibfield  {author} {\bibinfo {author} {\bibfnamefont {W.}~\bibnamefont
  {Lohstroh}}\ and\ \bibinfo {author} {\bibfnamefont {Z.}~\bibnamefont
  {Evenson}},\ }\bibfield  {title} {\enquote {\bibinfo {title} {Toftof: Cold
  neutron time-of-flight spectrometer},}\ }\href
  {https://doi.org/http://dx.doi.org/10.17815/jlsrf-1-40} {\bibfield  {journal}
  {\bibinfo  {journal} {Journal of large-scale research facilities}\ }\textbf
  {\bibinfo {volume} {1}},\ \bibinfo {pages} {A15} (\bibinfo {year}
  {2015})}\BibitemShut {NoStop}%
\bibitem [{\citenamefont {Copley}\ and\ \citenamefont {Cook}(2003)}]{DCS}%
  \BibitemOpen
  \bibfield  {author} {\bibinfo {author} {\bibfnamefont {J.}~\bibnamefont
  {Copley}}\ and\ \bibinfo {author} {\bibfnamefont {J.}~\bibnamefont {Cook}},\
  }\bibfield  {title} {\enquote {\bibinfo {title} {The disk chopper
  spectrometer at nist: a new instrument for quasielastic neutron scattering
  studies},}\ }\href
  {https://doi.org/https://doi.org/10.1016/S0301-0104(03)00124-1} {\bibfield
  {journal} {\bibinfo  {journal} {Chemical Physics}\ }\textbf {\bibinfo
  {volume} {292}},\ \bibinfo {pages} {477 -- 485} (\bibinfo {year} {2003})},\
  \bibinfo {note} {quasielastic Neutron Scattering of Structural Dynamics in
  Condensed Matter}\BibitemShut {NoStop}%
\bibitem [{\citenamefont {Bewley}, \citenamefont {Taylor},\ and\ \citenamefont
  {Bennington.}(2011)}]{BEWLEY_LET}%
  \BibitemOpen
  \bibfield  {author} {\bibinfo {author} {\bibfnamefont {R.}~\bibnamefont
  {Bewley}}, \bibinfo {author} {\bibfnamefont {J.}~\bibnamefont {Taylor}},\
  and\ \bibinfo {author} {\bibfnamefont {S.}~\bibnamefont {Bennington.}},\
  }\bibfield  {title} {\enquote {\bibinfo {title} {Let, a cold neutron
  multi-disk chopper spectrometer at isis},}\ }\href
  {https://doi.org/https://doi.org/10.1016/j.nima.2011.01.173} {\bibfield
  {journal} {\bibinfo  {journal} {Nuclear Instruments and Methods in Physics
  Research Section A: Accelerators, Spectrometers, Detectors and Associated
  Equipment}\ }\textbf {\bibinfo {volume} {637}},\ \bibinfo {pages} {128 --
  134} (\bibinfo {year} {2011})}\BibitemShut {NoStop}%
\bibitem [{\citenamefont {G.~Ehlers}\ and\ \citenamefont
  {Sokol}(2011)}]{EHLERS_CNCS}%
  \BibitemOpen
  \bibfield  {author} {\bibinfo {author} {\bibfnamefont {J.~N. E.~I.}\
  \bibnamefont {G.~Ehlers}, \bibfnamefont {A.~Podlesnyak}}\ and\ \bibinfo
  {author} {\bibfnamefont {P.}~\bibnamefont {Sokol}},\ }\bibfield  {title}
  {\enquote {\bibinfo {title} {The new cold neutron chopper spectrometer at the
  spallation neutron source: Design and performance},}\ }\href
  {https://doi.org/https://doi.org/10.1063/1.3626935} {\bibfield  {journal}
  {\bibinfo  {journal} {Review of Scientific Instruments 82, 085108 (2011)}\
  }\textbf {\bibinfo {volume} {637}},\ \bibinfo {pages} {128 -- 134} (\bibinfo
  {year} {2011})}\BibitemShut {NoStop}%
\bibitem [{\citenamefont {Nakajima}\ \emph {et~al.}(2011)\citenamefont
  {Nakajima}, \citenamefont {Ohira-Kawamura}, \citenamefont {Kikuchi},
  \citenamefont {Nakamura}, \citenamefont {Kajimoto}, \citenamefont {Inamura},
  \citenamefont {Takahashi}, \citenamefont {Aizawa}, \citenamefont {Suzuya},
  \citenamefont {Shibata}, \citenamefont {Nakatani}, \citenamefont {Soyama},
  \citenamefont {Maruyama}, \citenamefont {Tanaka}, \citenamefont {Kambara},
  \citenamefont {Iwahashi}, \citenamefont {Itoh}, \citenamefont {Osakabe},
  \citenamefont {Wakimoto}, \citenamefont {Kakurai}, \citenamefont {Maekawa},
  \citenamefont {Harada}, \citenamefont {Oikawa}, \citenamefont {E.~Lechner},
  \citenamefont {Mezei},\ and\ \citenamefont {Arai}}]{AMATERAS_Kenji}%
  \BibitemOpen
  \bibfield  {author} {\bibinfo {author} {\bibfnamefont {K.}~\bibnamefont
  {Nakajima}}, \bibinfo {author} {\bibfnamefont {S.}~\bibnamefont
  {Ohira-Kawamura}}, \bibinfo {author} {\bibfnamefont {T.}~\bibnamefont
  {Kikuchi}}, \bibinfo {author} {\bibfnamefont {M.}~\bibnamefont {Nakamura}},
  \bibinfo {author} {\bibfnamefont {R.}~\bibnamefont {Kajimoto}}, \bibinfo
  {author} {\bibfnamefont {Y.}~\bibnamefont {Inamura}}, \bibinfo {author}
  {\bibfnamefont {N.}~\bibnamefont {Takahashi}}, \bibinfo {author}
  {\bibfnamefont {K.}~\bibnamefont {Aizawa}}, \bibinfo {author} {\bibfnamefont
  {K.}~\bibnamefont {Suzuya}}, \bibinfo {author} {\bibfnamefont
  {K.}~\bibnamefont {Shibata}}, \bibinfo {author} {\bibfnamefont
  {T.}~\bibnamefont {Nakatani}}, \bibinfo {author} {\bibfnamefont
  {K.}~\bibnamefont {Soyama}}, \bibinfo {author} {\bibfnamefont
  {R.}~\bibnamefont {Maruyama}}, \bibinfo {author} {\bibfnamefont
  {H.}~\bibnamefont {Tanaka}}, \bibinfo {author} {\bibfnamefont
  {W.}~\bibnamefont {Kambara}}, \bibinfo {author} {\bibfnamefont
  {T.}~\bibnamefont {Iwahashi}}, \bibinfo {author} {\bibfnamefont
  {Y.}~\bibnamefont {Itoh}}, \bibinfo {author} {\bibfnamefont {T.}~\bibnamefont
  {Osakabe}}, \bibinfo {author} {\bibfnamefont {S.}~\bibnamefont {Wakimoto}},
  \bibinfo {author} {\bibfnamefont {K.}~\bibnamefont {Kakurai}}, \bibinfo
  {author} {\bibfnamefont {F.}~\bibnamefont {Maekawa}}, \bibinfo {author}
  {\bibfnamefont {M.}~\bibnamefont {Harada}}, \bibinfo {author} {\bibfnamefont
  {K.}~\bibnamefont {Oikawa}}, \bibinfo {author} {\bibfnamefont
  {R.}~\bibnamefont {E.~Lechner}}, \bibinfo {author} {\bibfnamefont
  {F.}~\bibnamefont {Mezei}},\ and\ \bibinfo {author} {\bibfnamefont
  {M.}~\bibnamefont {Arai}},\ }\bibfield  {title} {\enquote {\bibinfo {title}
  {Amateras: A cold-neutron disk chopper spectrometer},}\ }\href
  {https://doi.org/10.1143/JPSJS.80SB.SB028} {\bibfield  {journal} {\bibinfo
  {journal} {Journal of the Physical Society of Japan}\ }\textbf {\bibinfo
  {volume} {80}},\ \bibinfo {pages} {SB028} (\bibinfo {year}
  {2011})}\BibitemShut {NoStop}%
\bibitem [{\citenamefont {Andersen}\ \emph {et~al.}(2020)\citenamefont
  {Andersen}, \citenamefont {Argyriou}, \citenamefont {Jackson}, \citenamefont
  {Houston}, \citenamefont {Henry},\ and\ \citenamefont {etc
  ..}}]{ANDERSEN_ESS_Paper}%
  \BibitemOpen
  \bibfield  {author} {\bibinfo {author} {\bibfnamefont {K.}~\bibnamefont
  {Andersen}}, \bibinfo {author} {\bibfnamefont {D.}~\bibnamefont {Argyriou}},
  \bibinfo {author} {\bibfnamefont {A.}~\bibnamefont {Jackson}}, \bibinfo
  {author} {\bibfnamefont {J.}~\bibnamefont {Houston}}, \bibinfo {author}
  {\bibfnamefont {P.}~\bibnamefont {Henry}},\ and\ \bibinfo {author}
  {\bibfnamefont {P.~D.}\ \bibnamefont {etc ..}},\ }\bibfield  {title}
  {\enquote {\bibinfo {title} {The instrument suite of the european spallation
  source},}\ }\href
  {https://doi.org/https://doi.org/10.1016/j.nima.2020.163402} {\bibfield
  {journal} {\bibinfo  {journal} {Nuclear Instruments and Methods in Physics
  Research Section A: Accelerators, Spectrometers, Detectors and Associated
  Equipment}\ }\textbf {\bibinfo {volume} {957}},\ \bibinfo {pages} {163402}
  (\bibinfo {year} {2020})}\BibitemShut {NoStop}%
\bibitem [{\citenamefont {Hempelmann}(2001)}]{QENS_Hempelmann}%
  \BibitemOpen
  \bibfield  {author} {\bibinfo {author} {\bibfnamefont {R.}~\bibnamefont
  {Hempelmann}},\ }\href@noop {} {\emph {\bibinfo {title} {Quasielastic Neutron
  Scattering and Solid State Diffusion}}}\ (\bibinfo  {publisher} {Oxford
  University Press},\ \bibinfo {year} {2001})\BibitemShut {NoStop}%
\bibitem [{\citenamefont {Cai}\ \emph {et~al.}(2019)\citenamefont {Cai},
  \citenamefont {Wilson}, \citenamefont {Beare}, \citenamefont {Lygouras},
  \citenamefont {Thomas}, \citenamefont {Yahne}, \citenamefont {Ross},
  \citenamefont {Taddei}, \citenamefont {Sala}, \citenamefont {Dabkowska},
  \citenamefont {Aczel},\ and\ \citenamefont {Luke}}]{Cai_Er3Ga5O12}%
  \BibitemOpen
  \bibfield  {author} {\bibinfo {author} {\bibfnamefont {Y.}~\bibnamefont
  {Cai}}, \bibinfo {author} {\bibfnamefont {M.~N.}\ \bibnamefont {Wilson}},
  \bibinfo {author} {\bibfnamefont {J.}~\bibnamefont {Beare}}, \bibinfo
  {author} {\bibfnamefont {C.}~\bibnamefont {Lygouras}}, \bibinfo {author}
  {\bibfnamefont {G.}~\bibnamefont {Thomas}}, \bibinfo {author} {\bibfnamefont
  {D.~R.}\ \bibnamefont {Yahne}}, \bibinfo {author} {\bibfnamefont
  {K.}~\bibnamefont {Ross}}, \bibinfo {author} {\bibfnamefont {K.~M.}\
  \bibnamefont {Taddei}}, \bibinfo {author} {\bibfnamefont {G.}~\bibnamefont
  {Sala}}, \bibinfo {author} {\bibfnamefont {H.~A.}\ \bibnamefont {Dabkowska}},
  \bibinfo {author} {\bibfnamefont {A.~A.}\ \bibnamefont {Aczel}},\ and\
  \bibinfo {author} {\bibfnamefont {G.~M.}\ \bibnamefont {Luke}},\ }\bibfield
  {title} {\enquote {\bibinfo {title} {Crystal fields and magnetic structure of
  the ising antiferromagnet
  ${\mathrm{er}}_{3}{\mathrm{ga}}_{5}{\mathrm{o}}_{12}$},}\ }\href
  {https://doi.org/10.1103/PhysRevB.100.184415} {\bibfield  {journal} {\bibinfo
   {journal} {Phys. Rev. B}\ }\textbf {\bibinfo {volume} {100}},\ \bibinfo
  {pages} {184415} (\bibinfo {year} {2019})}\BibitemShut {NoStop}%
\bibitem [{\citenamefont {Zub{\'{a}}{\v{c}}}, \citenamefont {Javorsk{\'{y}}},\
  and\ \citenamefont {F{\aa}k}(2018)}]{Zub_2018_CEF}%
  \BibitemOpen
  \bibfield  {author} {\bibinfo {author} {\bibfnamefont {J.}~\bibnamefont
  {Zub{\'{a}}{\v{c}}}}, \bibinfo {author} {\bibfnamefont {P.}~\bibnamefont
  {Javorsk{\'{y}}}},\ and\ \bibinfo {author} {\bibfnamefont {B.}~\bibnamefont
  {F{\aa}k}},\ }\bibfield  {title} {\enquote {\bibinfo {title} {Crystal field
  in {NdPd}5al2 investigated by inelastic neutron scattering},}\ }\href
  {https://doi.org/10.1088/1361-648x/aac408} {\bibfield  {journal} {\bibinfo
  {journal} {Journal of Physics: Condensed Matter}\ }\textbf {\bibinfo {volume}
  {30}},\ \bibinfo {pages} {255801} (\bibinfo {year} {2018})}\BibitemShut
  {NoStop}%
\bibitem [{\citenamefont {Kitaev}(2006)}]{Kitaev}%
  \BibitemOpen
  \bibfield  {author} {\bibinfo {author} {\bibfnamefont {A.}~\bibnamefont
  {Kitaev}},\ }\bibfield  {title} {\enquote {\bibinfo {title} {Anyons in an
  exactly solved model and beyond},}\ }\href
  {https://doi.org/https://doi.org/10.1016/j.aop.2005.10.005} {\bibfield
  {journal} {\bibinfo  {journal} {Annals of Physics}\ }\textbf {\bibinfo
  {volume} {321}},\ \bibinfo {pages} {2 -- 111} (\bibinfo {year} {2006})},\
  \bibinfo {note} {january Special Issue}\BibitemShut {NoStop}%
\bibitem [{\citenamefont {Nayak}\ \emph {et~al.}(2008)\citenamefont {Nayak},
  \citenamefont {Simon}, \citenamefont {Stern}, \citenamefont {Freedman},\ and\
  \citenamefont {Das~Sarma}}]{RevModPhys_SpinLiquids_Nayak}%
  \BibitemOpen
  \bibfield  {author} {\bibinfo {author} {\bibfnamefont {C.}~\bibnamefont
  {Nayak}}, \bibinfo {author} {\bibfnamefont {S.~H.}\ \bibnamefont {Simon}},
  \bibinfo {author} {\bibfnamefont {A.}~\bibnamefont {Stern}}, \bibinfo
  {author} {\bibfnamefont {M.}~\bibnamefont {Freedman}},\ and\ \bibinfo
  {author} {\bibfnamefont {S.}~\bibnamefont {Das~Sarma}},\ }\bibfield  {title}
  {\enquote {\bibinfo {title} {Non-abelian anyons and topological quantum
  computation},}\ }\href {https://doi.org/10.1103/RevModPhys.80.1083}
  {\bibfield  {journal} {\bibinfo  {journal} {Rev. Mod. Phys.}\ }\textbf
  {\bibinfo {volume} {80}},\ \bibinfo {pages} {1083--1159} (\bibinfo {year}
  {2008})}\BibitemShut {NoStop}%
\bibitem [{\citenamefont {Banerjee}\ \emph {et~al.}(2017)\citenamefont
  {Banerjee}, \citenamefont {Yan}, \citenamefont {Knolle}, \citenamefont
  {Bridges}, \citenamefont {Stone}, \citenamefont {Lumsden}, \citenamefont
  {Mandrus}, \citenamefont {Tennant}, \citenamefont {Moessner},\ and\
  \citenamefont {Nagler}}]{Banerjee_quantumspinliquid}%
  \BibitemOpen
  \bibfield  {author} {\bibinfo {author} {\bibfnamefont {A.}~\bibnamefont
  {Banerjee}}, \bibinfo {author} {\bibfnamefont {J.}~\bibnamefont {Yan}},
  \bibinfo {author} {\bibfnamefont {J.}~\bibnamefont {Knolle}}, \bibinfo
  {author} {\bibfnamefont {C.~A.}\ \bibnamefont {Bridges}}, \bibinfo {author}
  {\bibfnamefont {M.~B.}\ \bibnamefont {Stone}}, \bibinfo {author}
  {\bibfnamefont {M.~D.}\ \bibnamefont {Lumsden}}, \bibinfo {author}
  {\bibfnamefont {D.~G.}\ \bibnamefont {Mandrus}}, \bibinfo {author}
  {\bibfnamefont {D.~A.}\ \bibnamefont {Tennant}}, \bibinfo {author}
  {\bibfnamefont {R.}~\bibnamefont {Moessner}},\ and\ \bibinfo {author}
  {\bibfnamefont {S.~E.}\ \bibnamefont {Nagler}},\ }\bibfield  {title}
  {\enquote {\bibinfo {title} {Neutron scattering in the proximate quantum spin
  liquid Œ±-rucl3},}\ }\href {https://doi.org/10.1126/science.aah6015}
  {\bibfield  {journal} {\bibinfo  {journal} {Science}\ }\textbf {\bibinfo
  {volume} {356}},\ \bibinfo {pages} {1055--1059} (\bibinfo {year} {2017})},\
  \Eprint
  {https://arxiv.org/abs/https://science.sciencemag.org/content/356/6342/1055.full.pdf}
  {https://science.sciencemag.org/content/356/6342/1055.full.pdf} \BibitemShut
  {NoStop}%
\bibitem [{\citenamefont {Pieper}(2009)}]{Pieper_TimeDependent}%
  \BibitemOpen
  \bibfield  {author} {\bibinfo {author} {\bibfnamefont {R.~G.}\ \bibnamefont
  {Pieper}, \bibfnamefont {J.}},\ }\bibfield  {title} {\enquote {\bibinfo
  {title} {Protein dynamics investigated by neutron scattering},}\ }\href
  {https://doi.org/10.1007/s11120-009-9480-9} {\bibfield  {journal} {\bibinfo
  {journal} {Photosynthesis Research}\ }\textbf {\bibinfo {volume} {102}},\
  \bibinfo {pages} {281} (\bibinfo {year} {2009})}\BibitemShut {NoStop}%
\bibitem [{\citenamefont {Mezei}(1988)}]{Mezei1988}%
  \BibitemOpen
  \bibfield  {author} {\bibinfo {author} {\bibfnamefont {F.}~\bibnamefont
  {Mezei}},\ }\bibfield  {title} {\enquote {\bibinfo {title} {Multilayer
  neutron optical devices: Physics, fabrication, and applications of
  multilayered structures. springer, boston},}\ }\href@noop {} {\  (\bibinfo
  {year} {1988})}\BibitemShut {NoStop}%
\bibitem [{\citenamefont {Ehlers}()}]{Priv_Comm_PersonG_Ehlers}%
  \BibitemOpen
  \bibfield  {author} {\bibinfo {author} {\bibfnamefont {G.}~\bibnamefont
  {Ehlers}},\ }\bibfield  {title} {\enquote {\bibinfo {title}
  {Private-communication},}\ }\href {https://doi.org/{2014}} {\
  {2014}}\BibitemShut {NoStop}%
\bibitem [{\citenamefont {Mezei}\ and\ \citenamefont
  {Russina}(2000)}]{MEZEI2000318}%
  \BibitemOpen
  \bibfield  {author} {\bibinfo {author} {\bibfnamefont {F.}~\bibnamefont
  {Mezei}}\ and\ \bibinfo {author} {\bibfnamefont {M.}~\bibnamefont
  {Russina}},\ }\bibfield  {title} {\enquote {\bibinfo {title} {Neutron beam
  extraction and delivery at spallation neutron sources},}\ }\href
  {https://doi.org/https://doi.org/10.1016/S0921-4526(00)00323-9} {\bibfield
  {journal} {\bibinfo  {journal} {Physica B: Condensed Matter}\ }\textbf
  {\bibinfo {volume} {283}},\ \bibinfo {pages} {318 -- 322} (\bibinfo {year}
  {2000})}\BibitemShut {NoStop}%
\bibitem [{\citenamefont {Kajimoto}\ \emph {et~al.}(2008)\citenamefont
  {Kajimoto}, \citenamefont {Nakajima}, \citenamefont {Nakamura}, \citenamefont
  {Osakabe}, \citenamefont {Sato},\ and\ \citenamefont
  {Arai}}]{AmaterasCurvedGuides}%
  \BibitemOpen
  \bibfield  {author} {\bibinfo {author} {\bibfnamefont {R.}~\bibnamefont
  {Kajimoto}}, \bibinfo {author} {\bibfnamefont {K.}~\bibnamefont {Nakajima}},
  \bibinfo {author} {\bibfnamefont {M.}~\bibnamefont {Nakamura}}, \bibinfo
  {author} {\bibfnamefont {T.}~\bibnamefont {Osakabe}}, \bibinfo {author}
  {\bibfnamefont {T.~J.}\ \bibnamefont {Sato}},\ and\ \bibinfo {author}
  {\bibfnamefont {M.}~\bibnamefont {Arai}},\ }\bibfield  {title} {\enquote
  {\bibinfo {title} {Curved neutron guide of the cold neutron disk-chopper
  spectrometer at j-parc},}\ }\href {https://doi.org/10.1080/10238160902819346}
  {\bibfield  {journal} {\bibinfo  {journal} {Journal of Neutron Research}\
  }\textbf {\bibinfo {volume} {16}},\ \bibinfo {pages} {81--86} (\bibinfo
  {year} {2008})}\BibitemShut {NoStop}%
\bibitem [{\citenamefont {{Willendrup, Peter Kjær}}\ and\ \citenamefont
  {{Lefmann, Kim.}}(2020)}]{McStas}%
  \BibitemOpen
  \bibfield  {author} {\bibinfo {author} {\bibnamefont {{Willendrup, Peter
  Kjær}}}\ and\ \bibinfo {author} {\bibnamefont {{Lefmann, Kim.}}},\
  }\bibfield  {title} {\enquote {\bibinfo {title} {Introduction, use, and basic
  principles for ray-tracing simulations},}\ }\href
  {https://doi.org/10.3233/JNR-190108} {\bibfield  {journal} {\bibinfo
  {journal} {Journal of Neutron Research}\ }\textbf {\bibinfo {volume} {22}},\
  \bibinfo {pages} {1--16} (\bibinfo {year} {2020})}\BibitemShut {NoStop}%
\bibitem [{\citenamefont {Stewart}\ \emph {et~al.}(2009)\citenamefont
  {Stewart}, \citenamefont {Deen}, \citenamefont {Andersen}, \citenamefont
  {Schober}, \citenamefont {Barth{\'{e}}l{\'{e}}my}, \citenamefont {Hillier},
  \citenamefont {Murani}, \citenamefont {Hayes},\ and\ \citenamefont
  {Lindenau}}]{Stewart_D7}%
  \BibitemOpen
  \bibfield  {author} {\bibinfo {author} {\bibfnamefont {J.~R.}\ \bibnamefont
  {Stewart}}, \bibinfo {author} {\bibfnamefont {P.~P.}\ \bibnamefont {Deen}},
  \bibinfo {author} {\bibfnamefont {K.~H.}\ \bibnamefont {Andersen}}, \bibinfo
  {author} {\bibfnamefont {H.}~\bibnamefont {Schober}}, \bibinfo {author}
  {\bibfnamefont {J.-F.}\ \bibnamefont {Barth{\'{e}}l{\'{e}}my}}, \bibinfo
  {author} {\bibfnamefont {J.~M.}\ \bibnamefont {Hillier}}, \bibinfo {author}
  {\bibfnamefont {A.~P.}\ \bibnamefont {Murani}}, \bibinfo {author}
  {\bibfnamefont {T.}~\bibnamefont {Hayes}},\ and\ \bibinfo {author}
  {\bibfnamefont {B.}~\bibnamefont {Lindenau}},\ }\bibfield  {title} {\enquote
  {\bibinfo {title} {{Disordered materials studied using neutron polarization
  analysis on the multi-detector spectrometer, D7}},}\ }\href
  {https://doi.org/10.1107/S0021889808039162} {\bibfield  {journal} {\bibinfo
  {journal} {Journal of Applied Crystallography}\ }\textbf {\bibinfo {volume}
  {42}},\ \bibinfo {pages} {69--84} (\bibinfo {year} {2009})}\BibitemShut
  {NoStop}%
\bibitem [{\citenamefont {Anastasopoulos}\ \emph {et~al.}(2017)\citenamefont
  {Anastasopoulos}, \citenamefont {Bebb}, \citenamefont {Berry}, \citenamefont
  {Birch}, \citenamefont {Bry{\'{s}}}, \citenamefont {Buffet}, \citenamefont
  {Clergeau}, \citenamefont {Deen}, \citenamefont {Ehlers}, \citenamefont {van
  Esch}, \citenamefont {Everett}, \citenamefont {Guerard}, \citenamefont
  {Hall-Wilton}, \citenamefont {Herwig}, \citenamefont {Hultman}, \citenamefont
  {H√∂glund}, \citenamefont {Iruretagoiena}, \citenamefont {Issa},
  \citenamefont {Jensen}, \citenamefont {Khaplanov}, \citenamefont {Kirstein},
  \citenamefont {Higuera}, \citenamefont {Piscitelli}, \citenamefont
  {Robinson}, \citenamefont {Schmidt},\ and\ \citenamefont
  {Stefanescu}}]{MG_Anastasopoulos_2017}%
  \BibitemOpen
  \bibfield  {author} {\bibinfo {author} {\bibfnamefont {M.}~\bibnamefont
  {Anastasopoulos}}, \bibinfo {author} {\bibfnamefont {R.}~\bibnamefont
  {Bebb}}, \bibinfo {author} {\bibfnamefont {K.}~\bibnamefont {Berry}},
  \bibinfo {author} {\bibfnamefont {J.}~\bibnamefont {Birch}}, \bibinfo
  {author} {\bibfnamefont {T.}~\bibnamefont {Bry{\'{s}}}}, \bibinfo {author}
  {\bibfnamefont {J.-C.}\ \bibnamefont {Buffet}}, \bibinfo {author}
  {\bibfnamefont {J.-F.}\ \bibnamefont {Clergeau}}, \bibinfo {author}
  {\bibfnamefont {P.}~\bibnamefont {Deen}}, \bibinfo {author} {\bibfnamefont
  {G.}~\bibnamefont {Ehlers}}, \bibinfo {author} {\bibfnamefont
  {P.}~\bibnamefont {van Esch}}, \bibinfo {author} {\bibfnamefont
  {S.}~\bibnamefont {Everett}}, \bibinfo {author} {\bibfnamefont
  {B.}~\bibnamefont {Guerard}}, \bibinfo {author} {\bibfnamefont
  {R.}~\bibnamefont {Hall-Wilton}}, \bibinfo {author} {\bibfnamefont
  {K.}~\bibnamefont {Herwig}}, \bibinfo {author} {\bibfnamefont
  {L.}~\bibnamefont {Hultman}}, \bibinfo {author} {\bibfnamefont
  {C.}~\bibnamefont {H√∂glund}}, \bibinfo {author} {\bibfnamefont
  {I.}~\bibnamefont {Iruretagoiena}}, \bibinfo {author} {\bibfnamefont
  {F.}~\bibnamefont {Issa}}, \bibinfo {author} {\bibfnamefont {J.}~\bibnamefont
  {Jensen}}, \bibinfo {author} {\bibfnamefont {A.}~\bibnamefont {Khaplanov}},
  \bibinfo {author} {\bibfnamefont {O.}~\bibnamefont {Kirstein}}, \bibinfo
  {author} {\bibfnamefont {I.~L.}\ \bibnamefont {Higuera}}, \bibinfo {author}
  {\bibfnamefont {F.}~\bibnamefont {Piscitelli}}, \bibinfo {author}
  {\bibfnamefont {L.}~\bibnamefont {Robinson}}, \bibinfo {author}
  {\bibfnamefont {S.}~\bibnamefont {Schmidt}},\ and\ \bibinfo {author}
  {\bibfnamefont {I.}~\bibnamefont {Stefanescu}},\ }\bibfield  {title}
  {\enquote {\bibinfo {title} {Multi-grid detector for neutron spectroscopy:
  results obtained on time-of-flight spectrometer {CNCS}},}\ }\href
  {https://doi.org/10.1088/1748-0221/12/04/p04030} {\bibfield  {journal}
  {\bibinfo  {journal} {Journal of Instrumentation}\ }\textbf {\bibinfo
  {volume} {12}},\ \bibinfo {pages} {P04030--P04030} (\bibinfo {year}
  {2017})}\BibitemShut {NoStop}%
\bibitem [{\citenamefont {{Zanini, Luca}}\ \emph {et~al.}(2020)\citenamefont
  {{Zanini, Luca}}, \citenamefont {{Klinkby, Esben}}, \citenamefont {{Mezei,
  Ferenc}},\ and\ \citenamefont {{Takibayev, Alan}}}]{ESS_Moderator}%
  \BibitemOpen
  \bibfield  {author} {\bibinfo {author} {\bibnamefont {{Zanini, Luca}}},
  \bibinfo {author} {\bibnamefont {{Klinkby, Esben}}}, \bibinfo {author}
  {\bibnamefont {{Mezei, Ferenc}}},\ and\ \bibinfo {author} {\bibnamefont
  {{Takibayev, Alan}}},\ }\bibfield  {title} {\enquote {\bibinfo {title}
  {Low-dimensional moderators at ess and compact neutron sources},}\ }\href
  {https://doi.org/10.1051/epjconf/202023104006} {\bibfield  {journal}
  {\bibinfo  {journal} {EPJ Web Conf.}\ }\textbf {\bibinfo {volume} {231}},\
  \bibinfo {pages} {04006} (\bibinfo {year} {2020})}\BibitemShut {NoStop}%
\end{thebibliography}%

\end{document}